\documentclass[]{aastex}
\usepackage{spr-astr-addons}
\usepackage{latexsym}
\usepackage{bm}

\usepackage{amsmath}

\RequirePackage{color}

\newcommand{\be}{\begin{equation}}
\newcommand{\ee}{\end{equation}}
\newcommand{\bn}{\begin{eqnarray}}
\newcommand{\en}{\end{eqnarray}}

\usepackage{bm}

\begin{document}

\title{Exponential Metric Fields}
\slugcomment{}
%% Running heads
\shorttitle{Exponential Metric Fields}
\shortauthors{Wytler C. dos Santos}

\author{Wytler Cordeiro dos Santos}
\affil{e-mail: wytler@fis.unb.br\\ Universidade de Bras\'ilia, Faculdade do
Gama, Bras\'ilia DF }
\email{wytler@fis.unb.br}

\begin{abstract}
The Laser Interferometer Space Antenna (LISA) mission will
use advanced technologies to achieve its science goals: the direct
detection of gravitational waves, the observation of signals from
compact (small and dense) stars as they spiral into black holes,
the study of the role of massive black holes in galaxy evolution,
the search for gravitational wave emission from the early
Universe. The gravitational red-shift, the advance of the perihelion of Mercury, 
deflection of light and the time delay of radar signals  are the 
classical tests in the first order of General Relativity (GR). However, LISA can possibly 
test Einstein's theories in the second order and perhaps, it will show
some particular feature of non-linearity of gravitational
interaction. In the present work we are seeking
a method to construct theoretical templates
that limit in the first order the tensorial structure of
some metric fields, thus the non-linear terms are given by
exponential functions of gravitational strength. 
The Newtonian limit obtained here, in the first order, is equivalent to GR.
\end{abstract}


%\section*{}
%\label{sec:intro}

\section{Introduction}

The extreme difficulties which arise if one tries to draw physically important conclusions from the basic assumptions
of Einstein's theory are mainly due to the non-linearity of the field equations. Moreover, the fact
that the spacetime topology is not given a priori, and the impossibility to integrate tensors over finite regions
cause difficulties unknown in other branches of mathematical physics.
Actually in this respect they are not so different from others fields,
for example the electromagnetic field, the scalar field, etc., by themselves obey linear equations in a
given spacetime, they form a non-linear system when their mutual interactions are taken into account.
The distinctive feature of the gravitational field is that it is self-interacting
(as the Yang-Mills field): it is non-linear even in the absence of other fields. 
This is because it defines the spacetime over which it propagates \citep{Hawking}.

Linearized gravity is any approximation to General Relativity
obtained from $g_{\mu\nu}=g_{\mu\nu}^{(0)}+h_{\mu\nu}$ (where
$g_{\mu\nu}^{(0)}$ is any curved background spacetime) in
Einstein's equation  and retaining only the terms linear in
$h_{\mu\nu}$ \citep{Wald}.
The weakness of the gravitational field means in the context of
general relativity that the spacetime is nearly flat. Small
gravitational perturbations in Minkowski space can be treated in
the simplest linearized version of General Relativity,
\begin{equation}
\label{linear}
g_{\mu\nu}=\eta_{\mu\nu}+h_{\mu\nu},
\end{equation}
as describing a theory of a
symmetric tensor field $h_{\mu\nu}$ propagating on at background spacetime.
This theory is Lorentz invariant in the sense of Special Relativity.
If one wants to obtain a solution of the non-linear equations, it
is necessary to employ an iterative method on approximate linear
equations whose solutions are shown to converge in a certain
neighbourhood of initial surface. It should be possible to avoid some of these difficulties of non-linearity by working in some spacetimes that shall be described in this paper. The proposal is that some metric fields can be separated into the parts carrying the dynamical information and those parts characterizing the coordinates system.
In this proposal,  terms of the coordinates system will have
tensorial structure limited  only in the first order. The
tensors that will describe $g_{\mu\nu}$ have linear behavior.
Naturally, there is a price that must be paid for the linear
tensors. The dynamic terms that carry the gravitational strength
have exponential structure. The principal idea came from the basic
principle that one should interpret (\ref{linear}) as separation
between pure mathematical and physical terms in metric field
tensor. In spite of the fact that $\eta_{\mu\nu}$ plays a key role
as empty flat and background spacetime of the Standard Model in
the description of fundamental interactions, this background
tensor metric $\eta_{\mu\nu}$ is an object wholly mathematical and
entirely geometrical, while $h_{\mu\nu}$ contains the physical
information. The strength of gravity is tied in the components of
$h_{\mu\nu}$. The proposal of this paper is a working hypothesis
to untie the strength of gravity from geometrical tensors. This
proposal is valid for a family of metric field tensors
$g_{\mu\nu}$, and some basic examples such Newtonian limit and
gravitational plane waves of low amplitudes are described.
%------------------------------------------------------------
%----------------------------------------------------------

This paper is outlined as follows: in Sec. II, we present the
basic mathematic concepts of the (quasi-)idempotent tensors that
compose the structure of metric fields approached in this work.
In Sec. III, we propose how to link the strength of gravity with
the tensors from Sec. II, then is defined a family of exponential
metrics. In Sec. IV, we present some examples of these exponential
metrics, such as: Yilmaz metric, circularly polarized wave and rotating
bodies. In Sec. V, we present exponential metrics (`adjoint metric
field') that are non-physical, but which help us to compute
Christoffel symbols, and consequently the curvature tensors, Ricci
tensors and determinant of metric field. In Sec. VI, we verify the
Newtonian limit and also we obtain gravitational waves. In
Sec. VII, we present a general conclusion.

We assume spacetime $({\cal M},{\bf g})$ to be a ${C^{\infty}4-}$dimen\-sional, globally hyperbolic,
pseudo-Riemannian manifold ${\cal M}$
with Lorentzian metric tensor ${\bf g}$ (whose components are $g_{\mu\nu}$) associated with the line element
$$ds^2=g_{\mu\nu}(x)dx^{\mu}dx^{\nu},$$
assumed to have  signature $(+---)$ \citep{Landau}. Lower case
Greek indices refer to coordinates on ${\cal M}$ and take the
values $0,1,2,3.$ The relation between the metric field
$g_{\mu\nu}$ and the material contents of spacetime is expressed
by Einstein's field equation,
\begin{equation}
\label{field equation}
R_{\mu\nu}-\frac{1}{2}g_{\mu\nu}R=\frac{8\pi G}{c^4}T_{\mu\nu},
\end{equation}
$T_{\mu\nu}$ being the stress-energy-momentum tensor, $R_{\mu\nu}$ the contract curvature tensor (Ricci tensor)
and $R$ its trace. In an empty region of spacetime we have $R_{\mu\nu}=0$,
such a region is called  vacuum field.

\section{Pure Mathematical terms of spacetime geometry}

The Minkowski flat spacetime $(\mathbb{R}^4,\bm\eta)$, where the
components of $\bm\eta$ are
$\eta_{\mu\nu}=\mbox{diag}(1,-1,-1,-1)$, is the simplest empty
spacetime and it is in fact the spacetime of Special Relativity. One
can obtain spaces locally identical to $(\mathbb{R}^4,\bm\eta)$
but with different (large scale) topology properties by
identifying points in $\mathbb{R}^4$ which are equivalent under a
discrete isometry without a fixed point. Any local Lorentz frames
is merely the statement that any curved space has the Minkowski
flat space `tangent' to it at any point. The Minkowski spacetime
is the universal covering space for all such derived spaces
\citep{Hawking}. In this sense it should be reasonable to one choose
$(\mathbb{R}^4,\bm\eta)$ as background spacetime and important
compound piece of some spacetimes. In fact, the most
straightforward approach to linear gravitation is realized in
Minkowski spacetime. The conformal structure of Minkowski space is
what one would regard as the `normal' behavior of a spacetime at
infinity.

The metric  tensor $\bm\eta$ from background Minkowski spacetime in any coordinates is
an object wholly mathematical and entirely geometrical. No strength of gravity is linked up with the
mathematical structure of $\bm\eta$. Then, we choose this metric tensor as a principal descriptive piece of some spacetimes
built below, and it does rise and lower the indices in the same way as in Special Relativity with $\eta_{\mu\nu}\eta^{\nu\alpha}={\delta_{\mu}}^{\alpha}$.
The aim of this paper is to obtain some spacetimes $({\cal M},{\bf g})$ that their non-linearity are less hard,
softer at least in a tensorial descriptive way.
Therefore, it is defined only a symmetric tensor $\bm\Upsilon$, which like $\bm\eta$
is an object wholly mathematical and entirely geometrical.
This (metric) tensor  $\bm\Upsilon$ that will be a piece of metric
${\bf g}$ can have non-static terms from spacetime in their
components, however in this approach, this tensor purely will not
have gravitational strength. The components $\Upsilon_{\mu\nu}$ of
tensor $\bm\Upsilon$ are raised and lowered by $\eta^{\mu\nu}$ and
$\eta_{\mu\nu}$,
\begin{eqnarray}
\eta^{\mu\nu}\Upsilon_{\nu\alpha}= {\Upsilon^{\mu}}_{\alpha}, \cr
\Upsilon_{\mu\nu}\eta^{\nu\alpha}= {\Upsilon_{\mu}}^{\alpha},\cr
\eta^{\mu\nu}\Upsilon_{\nu\alpha}\eta^{\alpha\beta}=\Upsilon^{\mu\beta}.
\end{eqnarray}
In this context it is adopted the point of view, that $\bm\Upsilon$
is a tensor on a background Minkowski spacetime, similar to
deviation $h_{\mu\nu}$ from linearized version of general
relativity (\ref{linear}). But instead one has the infinitesimal
condition for $|h_{\mu\nu}|\ll 1$, it is accepted that the
magnitude of $\Upsilon_{\mu\nu}$ can be equal to the magnitude of
empty flat spacetime ($|\Upsilon_{\mu\nu}|\approx |\eta_{\mu\nu}|
$). Moreover, it might be defined as an important mathematical
relationship among $\Upsilon_{\mu\nu}$ themselves,
\begin{eqnarray}
\label{ID}
\Upsilon_{\mu\nu}\Upsilon^{\nu\rho}=-2{\Upsilon_{\mu}}^{\rho}.
\end{eqnarray}
The above equation is an important argument to shape some
spacetimes that are described below. This equation  will improve
linearity in the tensorial sense. The contracting indices of tensor
$\bm\Upsilon$ by themselves  show that $\Upsilon_{\mu\nu}$
are (quasi-)idempotent elements 
($\bm\Upsilon\cdot \bm\Upsilon \propto \bm\Upsilon$; otherwise a factor $-2$ in the
operation), and the equation (\ref{ID}) improves at least to a linear tensorial template of
some tensor metric
fields.
It is possible to rise or lower indices of
$\Upsilon_{\mu\nu}$ operating themselves,
%\begin{small}
\begin{eqnarray}
\Upsilon^{\rho\mu}\Upsilon_{\mu\nu}\Upsilon^{\nu\sigma}=4\Upsilon^{\rho\sigma}.
\end{eqnarray}
%\end{small}
One can also verify the expression
${\Upsilon_{\mu}}^{\nu}{\Upsilon_{\nu\rho}}=-2\Upsilon_{\mu\rho}$.

The trace
of $\bm\Upsilon$ is obtained when $\rho=\mu$ in the expression (\ref{ID}),
\begin{eqnarray}
\label{ID traco}
\Upsilon_{\mu\nu}\Upsilon^{\nu\mu}=-2{\Upsilon_{\mu}}^{\mu},
\end{eqnarray}
and realize derivative calculation of trace
${\Upsilon_{\mu}}^{\mu}$ from (\ref{ID traco}) since $\eta_{\mu\nu}=\mbox{diag}(1,-1,-1,-1)$,
%\begin{small}
\begin{eqnarray}
\label{derivada traco1}
-2\partial_{\alpha}{\Upsilon_{\mu}}^{\mu} = 2 \Upsilon_{\mu\nu}\partial_{\alpha}\Upsilon^{\mu\nu},
\end{eqnarray}
%\end{small}
thus,
\begin{eqnarray}
\label{derivada traco2}
\Upsilon_{\mu\nu}\partial_{\alpha}\Upsilon^{\mu\nu}=\Upsilon^{\mu\nu}\partial_{\alpha}\Upsilon_{\mu\nu}=-\partial_{\alpha}{\Upsilon_{\mu}}^{\mu}.
\end{eqnarray}
Only both these tensors, $\bm\eta $ from background Minkowski
spacetime and $\bm\Upsilon$,  will become the mathematical and
geometrical basis to physical spacetimes described as follows.

\section{Strength of gravity terms tied in spacetime geometry}

A dimensionless parameter $\Phi$ characterizing the stren\-gth
of gravity at a spacetime point $\wp$ with coordinates $(t,{\bf
x})=(x^{\alpha})$ due to a gravitating source is the ratio 
of the potential energy, $m\varphi_{N}$ (due to this
source), to the inertial mass-energy $mc^2$ of a test body at $\wp$,
i.e.,
\begin{equation}
\Phi(x^{\alpha})=\frac{\varphi_{N}(x^{\alpha})}{c^2}.
\end{equation}
Here $\varphi_{N}(x^{\alpha})$ is the gravitational potential. For
a point source with mass $M$ in Newtonian gravity,
\begin{equation}
\Phi(x^{\alpha})=-\frac{GM}{c^2r},
\end{equation}
where $r$ is the distance to the source. So, for a nearly Newtonian system, we can use Newtonian
potential for $\varphi_{N}$.

To construct spacetimes with the basis $\bm\eta $ from background
Minkowski spacetime and $\bm\Upsilon$ defined in previous section,
it is proposed to tie the strength of gravity $\Phi$ to theses
tensors. While the tensor $\bm\Upsilon$ can be a function of the coordinates
$(t,{\bf x})=(x^{\alpha})$, the strength of gravity $\Phi$ will be
a function of the coordinates and also of the Newtonian gravitational
constant $G$  and of a parameter of mass $M$. Thus, I do propose:
\begin{equation}
{\bf g}(\Phi,x)=e^{\Phi}{\bm\eta}+\sinh(\Phi){\bm\Upsilon},\nonumber
\end{equation}
{with components}
\begin{equation}
\label{metricphi}
g_{\mu\nu}=e^{\Phi}\eta_{\mu\nu}+\sinh({\Phi})\Upsilon_{\mu\nu}.
\end{equation}
The inverse tensor is just
\begin{eqnarray}
\label{metricphiinverse}
g^{\mu\nu}=e^{-\Phi}\eta^{\mu\nu}-\sinh({\Phi})\Upsilon^{\mu\nu}.
\end{eqnarray}
If the tensor $\bm\Upsilon $ is diagonal, then the inverse tensor of ${\bf g}$ is
\begin{equation}
{\bf g}^{-1}(\Phi,x)={\bf g}(-\Phi,x)=e^{-\Phi}{\bm\eta}-\sinh(\Phi){\bm\Upsilon}.
\end{equation}
From the above definitions it follows that
$g_{\mu\nu}g^{\nu\alpha}={\delta_{\mu}}^{\alpha}$, in fact 
\begin{scriptsize}
\begin{eqnarray}
g_{\mu\nu}g^{\nu\alpha}=\left(e^{\Phi}\eta_{\mu\nu}+\sinh(\Phi)\Upsilon_{\mu\nu}\right)\left(e^{-\Phi}\eta^{\nu\alpha}-\sinh(\Phi)\Upsilon^{\nu\alpha}\right)\nonumber
\end{eqnarray}
\end{scriptsize}
where the rightside is 
\begin{scriptsize}
\begin{eqnarray}
{\delta_{\mu}}^{\alpha} +\sinh(\Phi)\left(e^{-\Phi}\Upsilon_{\mu\nu}\eta^{\nu\alpha}-e^{\Phi}\eta_{\mu\nu}\Upsilon^{\nu\alpha}-\sinh(\Phi)\Upsilon_{\mu\nu}\Upsilon^{\nu\alpha}\right)\nonumber
\end{eqnarray}
\end{scriptsize}
that is
\begin{scriptsize}
\begin{eqnarray}
{\delta_{\mu}}^{\alpha}+\sinh(\Phi)\left[-(e^{\Phi}-e^{-\Phi}){\Upsilon_{\mu}}^{\alpha}-\sinh(\Phi)\Upsilon_{\mu\nu}\Upsilon^{\nu\alpha}\right]&=&\cr
{\delta_{\mu}}^{\alpha}-\sinh^2(\Phi)\left(2{\Upsilon_{\mu}}^{\alpha}+\Upsilon_{\mu\nu}\Upsilon^{\nu\alpha}\right)
\end{eqnarray}
\end{scriptsize}
Using the expression (\ref{ID}) in the second term in the
parenthesis
($\Upsilon_{\mu\nu}\Upsilon^{\nu\alpha}=-2{\Upsilon_{\mu}}^{\alpha}$)
it results in,
\begin{eqnarray}
g_{\mu\nu}g^{\nu\alpha}&=&{\delta_{\mu}}^{\alpha}-\sinh^2(\Phi)\left[{2\Upsilon_{\mu}}^{\alpha}+(-2){\Upsilon_{\mu}}^{\alpha}\right],
\end{eqnarray}
finally one did prove this identity
$g_{\mu\nu}g^{\nu\alpha}={\delta_{\mu}}^{\alpha}$. Because of equation
(\ref{ID}), there are no quadratic values  naturally in
$\bm\Upsilon$ and  consequently the non-linearity of metric field
will be less complicated.

If one considers some spacetime $({\cal M},{\bf g})$ such as the same of 
${\bf g}$ from definition (\ref{metricphi}), one can
observe that the first term
$\gamma_{\mu\nu}=e^{\Phi}\eta_{\mu\nu}$ is a background spacetime
conformally flat (where the Weyl tensor vanishes). In the previous
section one has accepted that $|\Upsilon_{\mu\nu}|\approx
|\eta_{\mu\nu}|$, but now because of $\sinh{(\Phi)}$ (that can be
small) multiplying this tensor $\bm\Upsilon$, it can be understood
as (small) disturbance ${\cal S}_{\mu\nu}=\sinh(\Phi)\Upsilon_{\mu\nu}$, such as,
\begin{equation}
g_{\mu\nu}=\gamma_{\mu\nu}+{\cal S}_{\mu\nu},
\end{equation}
if one is
dealing with the Einstein's vacuum equation, the (small)
disturbance that represents the gravitational wave can be
separated from the conformally flat background spacetime
$\gamma_{\mu\nu}$ \citep{Birrell}.

\section{Some Examples}
\subsection{Yilmaz Metric}

As a first application of the metric field defined in previous
sections, let us take ${\Upsilon}_{\mu\nu} $ to be:
\begin{small}
\begin{eqnarray}
\label{weak} {\Upsilon}_{\mu\nu}=
\begin{pmatrix}0&0&0&0\cr
         0&2&0&0 \cr
         0&0&2&0  \cr
         0&0&0&2
\end{pmatrix},\hspace{0.1cm}{{\Upsilon}_{\mu}}^{\nu}=%\Upsilon_{\mu\alpha}\eta^{\alpha\nu}=
\begin{pmatrix}0&0&0&0\cr
         0&-2&0&0 \cr
         0&0&-2&0  \cr
         0&0&0&-2
\end{pmatrix}\nonumber
\end{eqnarray}
\end{small}
and
\begin{small}
\begin{eqnarray}
{\Upsilon}^{\mu\nu}=
\begin{pmatrix}0&0&0&0\cr
         0&2&0&0 \cr
         0&0&2&0  \cr
         0&0&0&2
\end{pmatrix}
\end{eqnarray}
\end{small}
that ${\Upsilon}_{\mu\nu}{\Upsilon}^{\nu\alpha}=-2{{\Upsilon}_{\mu}}^{\alpha}$, with trace given ${\Upsilon}_{\mu\nu}{\Upsilon}^{\nu\mu}=-2{{\Upsilon}_{\mu}}^{\mu}=-2{\bf Tr} {\bm\Upsilon}$, that ${\bf Tr} ({\bm\Upsilon})=-6$ .
Now we can display the tensor metric (\ref{metricphi}),
\begin{scriptsize}
\begin{eqnarray}
\label{metricphi2}
g_{\mu\nu}&=&e^{\Phi}\begin{pmatrix}1&0&0&0\cr
         0&-1&0&0 \cr
         0&0&-1&0  \cr
         0&0&0&-1
\end{pmatrix}+\sinh(\Phi)\begin{pmatrix}0&0&0&0\cr
         0&2&0&0 \cr
         0&0&2&0  \cr
         0&0&0&2
\end{pmatrix}\cr
   g_{\mu\nu} &=&
\begin{pmatrix}
         e^{\Phi}&0&0&0\cr
         0&-e^{-\Phi}&0&0 \cr
         0&0&-e^{-\Phi}&0  \cr
         0&0&0&-e^{-\Phi}
\end{pmatrix},
\end{eqnarray}
\end{scriptsize}
since $-e^{\Phi}+2\sinh(\Phi)=-e^{-\Phi}$. Then the line element is,
\be
\label{Yilmaz}
ds^2=e^{\Phi}c^2dt^2 - e^{-\Phi}(dx^2+dy^2+dz^2).
\ee
This metric field (\ref{metricphi2},\ref{Yilmaz}) has been proposed by Yilmaz \citep{Yilmaz1,Yilmaz2,Yilmaz3,Yilmaz4,Yilmaz5,Yilmaz6,Yilmaz7}. In the case of a mass singularity, $\Phi=-\dfrac{2GM}{c^2r}\ll 1$
we have  the far-field metric,
\begin{small}
\be
\label{metricphi3}
ds^2=(1-\frac{2GM}{c^2r})c^2dt^2-(1+\frac{2GM}{c^2r})(dx^2+dy^2+dz^2).
\ee
\end{small}
This is to be contrasted with the Schwarzschild (in General Relativity, GR) line element in isotropic coordinates \citep{Landau},
\begin{scriptsize}
\be
\label{isotropic}
ds^2=\left(\frac{1+\Phi/4}{1-\Phi/4}\right)^2c^2dt^2 - \left(1-\frac{\Phi}{4}\right)^4(dr^2+ r^2d\theta^2+r^2\sin^2\theta d\phi^2),
\ee
\end{scriptsize}
if we compare expansions with the line element of Yilmaz theory (\ref{Yilmaz}),\\
\indent Yilmaz:
\begin{scriptsize}
\begin{eqnarray}
g_{00}&=&1+\Phi+\frac{\Phi^2}{2}+\frac{\Phi^3}{6}+\cdots \hspace{0.5cm} g_{11}=-1+\Phi-\frac{\Phi^2}{2}+\cdots\nonumber
\end{eqnarray}
\end{scriptsize}
\indent GR:
\begin{scriptsize}
\begin{eqnarray}
g_{00}&=&1+\Phi+\frac{\Phi^2}{2}+\frac{3\Phi^3}{16}+\cdots \hspace{0.5cm} g_{11}=-1+\Phi-\frac{3\Phi^2}{8}-\cdots\,,\nonumber
\end{eqnarray}
\end{scriptsize}
$g_{11}$ coefficients differing only in the second order terms, 
while $g_{00}$ differing in the third order.
Both, Yilmaz and GR, give observational indistinguishable predictions for red-shift, light bending and perihelion advance, but the Yilmaz metric does not admit black holes. Citing this property and assuming that Yilmaz theory is correct, Clapp \citep{Clapp} has suggest that a significant component of quasar red-shift may be gravitational. Robertson \citep{Robertson1, Robertson2} has suggested that some neutron stars and black hole candidates may be like `Yilmaz stars'. Robertson argues that neutron star with mass $\sim 10M_{\odot}$ is found for Yilmaz metric while that an object of nuclear density greater than
$\sim 2.8M_{\odot}$ should be a black hole in Schwarzschild metric.
Ibison has tested Yilmaz theory by working out the corresponding Friedmann equations generated by assuming the Friedmann-Robertson-Walker cosmological metrics \citep{Ibison}. There are a series of claims and counter-claims involving Fackerell \citep{Fackerell,Yilmaz8,Alley1}, and also Misner and Wyss \citep{Wyss,Misner,Alley2} about Yilmaz theory.
At the present time both Yilmaz and
Schwarzschild solutions give results in agreement with
observation \citep{Rosen}.
However, it may be possible in the future, with LISA mission \citep{NASA,Baker}, to distinguish between Yilmaz and Schwarzschild.

\subsection{Circularly Polarized Wave}
Gravitational waves are one of the most important predictions of General Relativity.
Now we can try a solution of gravitational waves in $z$ direction,
\begin{scriptsize}
\begin{eqnarray}
 \label{wavephi}
{\Upsilon}_{\mu\nu}&=&
\begin{pmatrix}0&0&0&0\cr
         0&1+\cos{\zeta}&\sin{\zeta}&0 \cr
         0&\sin{\zeta}&1-\cos{\zeta}&0  \cr
         0&0&0&2
\end{pmatrix},\cr
{{\Upsilon}_{\mu}}^{\nu}&=&{\Upsilon}_{\mu\alpha}\eta^{\alpha\nu}=
\begin{pmatrix}0&0&0&0\cr
         0&-1-\cos{\zeta}&-\sin{\zeta}&0 \cr
         0&-\sin{\zeta}&-1+\cos{\zeta}&0  \cr
         0&0&0&-2
\end{pmatrix}
\end{eqnarray}
\end{scriptsize}
and
\begin{scriptsize}
\begin{equation}
\Upsilon^{\mu\nu}=\Upsilon_{\alpha\beta}\,\eta^{\alpha\mu}\eta^{\beta\nu}=
\begin{pmatrix}0&0&0&0\cr
         0&1+\cos{\zeta}&\sin{\zeta}&0 \cr
         0&\sin{\zeta}&1-\cos{\zeta}&0  \cr
         0&0&0&2
\end{pmatrix}
\end{equation}
\end{scriptsize}
so  ${\Upsilon}_{\mu\nu}{\Upsilon}^{\nu\alpha}=-2{{\Upsilon}_{\mu}}^{\alpha}$ is verified 
for the tensor above and ${\bf Tr} ({\bm\Upsilon})=-4$.
Then, if one chooses $\zeta=\omega t-kz$,  the metric field is a solution for 
a  gravitational plane wave
$g_{\mu\nu}=e^\Phi\eta_{\mu\nu}+\sinh(\Phi)\Upsilon_{\mu\nu} $,
%\begin{scriptsize}
\begin{eqnarray}
g_{\mu\nu}&=&
\begin{pmatrix}e^{\Phi}&0&0&0\cr
         0&-\cosh(\Phi)&0&0 \cr
         0&0&-\cosh(\Phi)&0  \cr
         0&0&0&-e^{\Phi}
\end{pmatrix}\cr
&+&\sinh(\Phi)
\begin{pmatrix}0&0&0&0\cr
         0&\cos{\zeta}&\sin{\zeta}&0 \cr
         0&\sin{\zeta}&-\cos{\zeta}&0  \cr
         0&0&0&2
\end{pmatrix}.
\end{eqnarray}
%\end{scriptsize}
Where  the first term can be the background spacetime
(asymptotically flat) and the second term is the disturbance in this
background or in other words the circularly polarized radiation
$h_{\mu\nu}^{TT}$ with amplitude $\sinh(\Phi)$.
The gravitational wave polarization is important from astrophysical and cosmological viewpoints. 
A binary system of two stars in circular orbit one around the other is expected to emit circularly polarized waves in the direction perpendicular 
to the plane of the orbit \citep{Schutz}.
Moreover, the Big Bang left behind an echo in the electromagnetic spectrum, the cosmic microwave background, but the Big Bang most 
likely also left cosmological gravitational waves that will be possible to observe with the help of LISA \citep{NASA,Baker}. 
Since cosmological gravitational waves propagate without significant interaction after they are produced, once detected they 
should provide a powerful tool for studying the early Universe at the time of gravitational wave generation \citep{Buonanno}.
Various mechanisms for cosmological gravitational wave generation have been proposed, and many of these state that the 
cosmological gravitational wave are circularly polarized. T. Kahniashvili et al \citep{Kahniashvili} argued that  helical 
turbulence produced during a first-order phase transition generated circularly
polarized cosmological gravitational waves. Other physicists  have said that the parity violation  due to the
gravitational Chern-Simons term in superstring theory can produce the primordial gravitational waves with circular polarization \citep{Lue,Choi,Alexander,Satoh,Saito}.

If we assume long distance from source, we can obtain plane wave solution with $\Phi\ll 1$ so that,
\begin{scriptsize}
\begin{eqnarray}
\label{wavephi2}
g_{\mu\nu}&=&(1+\Phi)\eta_{\mu\nu}+\Phi\Upsilon_{\mu\nu}\cr
&=&
\begin{pmatrix}1+\Phi&0&0&0\cr
         0&-1&0&0 \cr
         0&0&-1&0  \cr
         0&0&0&-1+\Phi
\end{pmatrix}\cr
& &+\Phi\begin{pmatrix}0&0&0&0\cr
         0&\cos{\zeta}&\sin{\zeta}&0 \cr
         0&\sin{\zeta}&-\cos{\zeta}&0  \cr
         0&0&0&0
\end{pmatrix}\nonumber
\end{eqnarray}
\end{scriptsize}
the second term is the circularly polarized radiation $h_{\mu\nu}^{TT}$. Because of the far distance from source, we have the presence of $g_{00}$ and $g_{33}$ as static terms, weakly pertubed  in these coordinates.

\subsection{Rotating Bodies}

The Kerr metric is important astrophysically since it is a good
approximation to the metric of a rotating star at large distances.
It is possible to obtain a kind Kerr metric from
$g_{\mu\nu}=e^{\Phi}\eta_{\mu\nu}+\sinh(\Phi)\Upsilon_{\mu\nu}$,
that in coordinates $(t,x,y,z)$ the tensor $\Upsilon_{\mu\nu}$ is:
\begin{tiny}
\begin{equation}
\label{UKerr}
%\Upsilon_{\mu\nu}=(-2)
\begin{pmatrix}
                            \cosh^2\Lambda & -\sinh\Lambda\cosh\Lambda \sin{\phi} & \sinh\Lambda\cosh\Lambda \cos{\phi} & 0\cr
                    -\sinh\Lambda\cosh\Lambda \sin{\phi} & \sinh^2\Lambda\sin^2\phi & -\sinh^2\Lambda\cos\phi \sin{\phi} & 0\cr
                    \sinh\Lambda\cosh\Lambda \cos{\phi}& -\sinh^2\Lambda\cos{\phi}\sin{\phi} & \sinh^2\Lambda\cos^2{\phi} & 0\cr
                                        0 & 0  &  0  & 0
\end{pmatrix}
\end{equation}
\end{tiny}
satisfying $\Upsilon_{\mu\nu}\Upsilon^{\nu\rho}=-2{\Upsilon_{\mu}}^{\rho}$ with ${\bf Tr}({\bm\Upsilon})=-2$. In this example we choose $\Upsilon_{33}=0$, but if the choice was $\Upsilon_{33}=2$, the above tensor still satisfy the algebra (\ref{ID}).
One can change coordinates $(t,x,y,z)$ to the Boyer-Lindquist  coordinates $(t,r,\theta,\phi)$,
with spatial part as flat space in ellipsoidal coordinates,
\begin{eqnarray}
\label{Elipse achatada}
t&=&t,\cr
x&=&\sqrt{r^2+a^2}\sin\theta\cos\phi,\cr
y&=&\sqrt{r^2+a^2}\sin\theta\sin\phi,\cr
z&=&r\cos\theta,
\end{eqnarray}
The Minkowski tensor metric related to them is:
\begin{scriptsize}
\begin{equation}
\label{eta elipsi}
\eta_{\mu\nu}=
\begin {pmatrix}
1&0&0&0\\\noalign{\medskip}0&-{\frac {{r}
^{2}+{a}^{2}  \cos ^2 \theta}{{r}^{2}+
{a}^{2}}}&0&0\\\noalign{\medskip}0&0&-{r}^{2}-{a}^{2}\cos^2\theta &0\\\noalign{\medskip}0&0&0&-
 \left( {r}^{2}+{a}^{2} \right) \sin^2 \theta
\end{pmatrix},
\end{equation}
\end{scriptsize}
we are assuming that the angle $\phi$ from tensor $\Upsilon_{\mu\nu}$ of (\ref{UKerr}) can be the same angle from transformations (\ref{Elipse achatada}). So, this coordinate transformations will become tensor (\ref{UKerr}) in:
\begin{tiny}
\begin{equation}
\label{UKerr2}
\Upsilon_{\mu\nu}=
(-2)\begin{pmatrix} \cosh^2 \Lambda &0&0&\sinh  \Lambda  \cosh  \Lambda R\sin \theta
\\\noalign{\medskip}0&0&0&0\\\noalign{\medskip}0&0&0&0
\\\noalign{\medskip}\sinh  \Lambda  \cosh  \Lambda R\sin \theta &0&0&
 \sinh^2\Lambda R^2   \sin^2\theta
\end{pmatrix}
\end{equation}
\end{tiny}
where $R=\sqrt {{r}^{2}+{a}^{2}}$.
At the appendices, it is verified with details that the above tensor obeys the
algebra (\ref{ID}) in the background Minkowski spacetime (\ref{eta elipsi}). 
Now we can choose a particular solution for this tensor choosing the geometric terms $\sinh\Lambda$ and $\cosh\Lambda$:
\begin{equation}
\label{definicao shL}
\sinh\Lambda=-\frac{a\sin\theta}{\rho} \hspace{0.5cm}\mbox{and} \hspace{0.5cm}\cosh\Lambda=\frac{\sqrt{r^2+a^2}}{\rho} ,
\end{equation}
with
\begin{equation}
\rho^2=r^2+a^2\cos^2\theta
\end{equation}
satisfying $\cosh^2\Lambda -\sinh^2\Lambda=1$.

The physical terms that contain the strength  of gravity
$e^{\Phi}$ and $\sinh{\Phi}$ can be: \footnote{In this paragraph
we assume $G=1$ and $c=1$.}
\begin{equation}
\label{Phi RB}
\Phi=\frac{Mr}{(r^2+a^2)}\ll 1,
\end{equation}
so that:
\begin{equation}
e^{\Phi}\approx 1+\Phi =1+ \frac{Mr}{(r^2+a^2)} \nonumber
\end{equation}
{and} 
\begin{equation}
\sinh(\Phi)\approx \Phi= \frac{Mr}{(r^2+a^2)}.\nonumber
\end{equation}
We can compute each term of metric field $g_{\mu\nu}$ 
(for more details see appendices),
\begin{small}
\begin{eqnarray}
g_{00}&=&\frac{\Delta-a^2\sin^2\theta}{\rho^2}+\frac{Mr}{(r^2+a^2)},\cr
g_{03}&=&g_{30}=\frac{2Mra\sin^2{\theta}}{\rho^2},\cr
g_{11}&=& -\frac{\rho^2}{\Delta}+ \frac{Mr\rho^2}{\Delta(r^2+a^2)},\cr
g_{22}&=&-\rho^2-\frac{Mr\rho^2}{r^2+a^2},\cr
g_{33}&=&\frac{-\sin^2\theta\left[\left(r^2+a^2\right)^2-\Delta a^2\sin^2\theta\right]}{\rho^2} - Mr\sin^2\theta,\nonumber
\end{eqnarray}
\end{small}
 where we use the definition ,
\begin{equation}
\Delta=r^2-2Mr+a^2.
\end{equation}
The metric tensor  $g_{\mu\nu}$ in the matrix form is:
\begin{small}
\begin{equation}
\begin{pmatrix}\frac{\Delta-a^2\sin^2\theta}{\rho^2}& 0 & 0 & \frac{2Mra\sin^2{\theta}}{\rho^2}\cr
0&-\frac{\rho^2}{\Delta}& 0 & 0  \cr
0 & 0 & -\rho^2 & 0 \cr
\frac{2Mra\sin^2{\theta}}{\rho^2}& 0 & 0 & \frac{-\sin^2\theta\left[\left(r^2+a^2\right)^2-\Delta a^2\sin^2\theta\right]}{\rho^2} 
\end{pmatrix}\nonumber
\end{equation}
\begin{equation}
+
\begin{pmatrix} \frac{Mr}{(r^2+a^2)}& 0 & 0 &0 \cr
0 & \frac{Mr\rho^2}{\Delta(r^2+a^2)}& 0 & 0  \cr
0 & 0 & -\frac{Mr\rho^2}{r^2+a^2} & 0 \cr
0 & 0 & 0 & - Mr\sin^2\theta
\end{pmatrix}.
\end{equation}
\end{small}
%\end{footnotesize}
The above first matrix is just `exact Kerr solution'. While the second matrix can be seen as deviation
or deformity from `exact solution'. Then, it makes sense to see this deviation as approximate solution of
\begin{equation}
g_{\mu\nu}=g_{\mu\nu}^{(0)}+h_{\mu\nu},
\end{equation}
where $g_{\mu\nu}^{(0)}$ is some known exact solution (that in
this case is Kerr solution) and $h_{\mu\nu}$ is the perturbation.
For Einstein's vacuum equation it is possible  to obtain explicitly
the vacuum perturbation equations from an arbitrary exact solution
\citep{Wald}.

Gravitational wave observations of extreme-mass-ratio-inspirals (EMRIs) by LISA will
provide unique evidence for the identity of the supermassive objects in galactic nuclei.
It is commonly assumed that these objects are indeed Kerr black holes.
K. Glampedakis and S. Babak argue that
from the observed signal, LISA will have the potential to prove (or disprove) this assumption,
by extracting the first few multipole moments of the spacetime outside these objects.
The possibility of discovering a non-Kerr object should be taken into account when
constructing waveform templates for LISA's data analysis tools. They provide a prescription for building a `quasi-Kerr' metric, that is a metric that slightly deviates
from Kerr, and present results on how this deviation impacts orbital motion and the emitted waveform \citep{Glampedakis}.

\subsection{Deformed Schwarzschild spacetime}
Another example can be given in spherical coordinates
$(t,r,\theta,\phi)$ where the components of metric tensor of  Minkowski flat spacetime is:\\ $\eta_{\mu\nu}=\mbox{diag}(1,-1,-r^2,-r^2\sin\theta)$,\\ with the respective inverse\\ $\eta^{\mu\nu}=\mbox{diag}(1,-1,-\frac{1}{r^2},-\frac{1}{r^2\sin\theta})$.
As we have seen before, let us describe a simpler (quasi-)idempotent tensor in spherical coordinates given by
\begin{equation}
\label{deformed S}
\Upsilon_{\mu\nu}=
\begin{pmatrix}
0&0&0&0\cr
0&2&0&0\cr
0&0&{r}^{2}&-{r}^{2}\sin\theta\cr
0&0&-{r}^{2}\sin\theta &{r}^{2} \sin^2 \theta
\end{pmatrix}
\end{equation}
with
\begin{equation}
{\Upsilon_{\mu}}^{\nu}=\Upsilon_{\mu\alpha}\eta^{\alpha\nu}=
\begin{pmatrix}
0&0&0&0\cr
0&-2&0&0\cr
0&0&-1&\frac{1}{\sin\theta}\cr
0&0&\sin\theta &-1
\end{pmatrix}
\end{equation}
{and} 
\begin{equation}
{\Upsilon^{\mu}}_{\nu}=\eta^{\mu\alpha}\Upsilon_{\alpha\nu}=
\begin{pmatrix}
0&0&0&0\cr
0&-2&0&0\cr
0&0&-1&{\sin\theta}\cr
0&0&\frac{1}{\sin\theta} &-1
\end{pmatrix}.
\end{equation}
Then ${\bf Tr}(\bm\Upsilon)={\Upsilon^{\mu}}_{\mu}=-4$.
One can show that $\Upsilon^{\mu\nu}$ given by $\eta^{\mu\alpha}\Upsilon_{\alpha\beta}\eta^{\beta\nu}$ is,
\begin{equation}
\Upsilon^{\mu\nu}=
\begin{pmatrix}
0&0&0&0\cr
0&2&0&0\cr
0&0&\frac{1}{r^2}&-{\frac {1}{{r}^{2}\sin \theta}}\cr
0&0&-{\frac {1}{{r}^{2}\sin \theta }}&{\frac {1}{{r}^{2} \sin^2\theta}}
\end{pmatrix}
\end{equation}
that satisfies the algebraic relation (\ref{ID}),
\begin{small}
\begin{eqnarray}
\Upsilon_{\mu\alpha}\Upsilon^{\alpha\nu}&=&
\begin{pmatrix}
0&0&0&0\cr
0&4&0&0\cr
0&0&2&-\frac{2}{\sin \theta}\cr
0&0&-2\,\sin \theta  &2
\end{pmatrix}\cr
&=&-2
\begin{pmatrix}
0&0&0&0\cr
0&-2&0&0\cr
0&0&-1&\frac{1}{\sin \theta}\cr
0&0&\sin \theta  &-1
\end {pmatrix}=-2{\Upsilon_{\mu}}^{\nu}.
\end{eqnarray}
\end{small}
From expression (\ref{metricphi}) we have the metric field $g_{\mu\nu}=e^{\Phi}\eta_{\mu\nu}+\sinh\Phi\Upsilon_{\mu\nu}$,
it follows that,
\begin{eqnarray}
g_{00}&=&e^{\Phi},\cr
g_{rr}&=&-\frac{1}{e^{\Phi}},\cr
g_{\theta\theta}&=&-r^2\cosh\Phi,\cr
g_{\phi\phi}&=&-r^2\sin^2\theta\cosh\Phi,\cr
g_{\theta\phi}&=&g_{\phi\theta}=-r^2\sin\theta\sinh\Phi,
\end{eqnarray}
in the case that $\Phi=-\frac{2GM}{c^2r}\ll 1$ with line element expanded in the first order of the strength of gravity,
\begin{scriptsize}
\begin{eqnarray}
ds^2 &=& \underbrace{\left(1-\frac{2GM}{c^2r}\right)c^2dt^2-\frac{dr^2}{\left(1-\frac{2GM}{c^2r}\right)}-r^2d\theta^2-r^2\sin^2\theta d\phi^2}_{\mbox{Schwarzschild spacetime}}\cr  &+& 4\left(\frac{GM}{c^2}\right) r \sin\theta d\theta d\phi,
\end{eqnarray}
\end{scriptsize}
This metric field is asymptotically flat, the components approach those of Minkowski spacetime in
spherical coordinates.
If only $g_{\theta\phi}$ would be vanished, we could obtain the Schwarzchild spacetime. 
However, it is necessary the terms 
$\Upsilon_{\theta\phi}= \Upsilon_{\phi\theta}= -r^2\sin\theta$ in the symmetric tensor 
of equation (\ref{deformed S}) for this tensor may obey the algebraic relation (\ref{ID}).
Nevertheless, it raises a question: what kind of gravitating source should distort a spherically symmetric spacetime?
The main purpose of this section is to
illustrate some (quasi-)idempotent tensors $\bm\Upsilon$ that must satisfy the algebraic relation (\ref{ID}).
There are many works with discussions about Yilmaz metric field and their sources.
However, this paper does not discuss the physics of gravitating sources of
circularly polarized wave, rotating bodies and deformed Schwarzschild spacetime purposed here.
In forthcoming work, it will be necessary to analyse principally the gravitating source
that distort the static Schwarzschild solution.

These examples suggest that the ${\bf Tr}({\bm\Upsilon})\in\mathbb{Z}$ are constant numbers.
Forward it will be necessary to calculate the derivative of  ${\bf Tr}({\bm\Upsilon})$ in any situations,
thus we shall assume that,
\begin{equation}
\label{Dtraca}
\partial_\alpha{\bf Tr}({\bm\Upsilon})=0
\end{equation}
in this paper.

\section{Adjoint Metric Field, Christoffel Symbols and determinant}
\subsection{Adjoint Metric Field}

One can see that spacetime (\ref{metricphi}) is asymptotically
flat. The Minkowski spacetime is the universal covering space for
all such derived spacetimes, and in this sense it is possible to
restore Minkowski  spacetime in (\ref{metricphi}) if one turns off
the gravitational strength in metric field, or  in other  words,
if $\Phi=0$ then $g_{\mu\nu}=\eta_{\mu\nu}$. However one can
construct a kind of spacetime with hyperbolic cosine instead of
hyperbolic sine that is not asymptotically flat,
\begin{equation}
\label{pseudometricphi}
\breve{g}_{\mu\nu}=e^{\Phi}\eta_{\mu\nu}+\cosh(\Phi)\Upsilon_{\mu\nu},
\end{equation}
with respective inverse,
\begin{equation}
\label{pseudometricphiinverse}
\breve{g}^{\mu\nu}=e^{-\Phi}\eta^{\mu\nu}+\cosh(\Phi)\Upsilon^{\mu\nu},
\end{equation}
and with $\breve{g}_{\mu\nu}\breve{g}^{\nu\alpha}={\delta_{\mu}}^{\alpha}$
since
$\Upsilon_{\mu\nu}\Upsilon^{\nu\alpha}=-2{\Upsilon_{\mu}}^{\alpha}$
from equation (\ref{ID}). A map between metrics field
(\ref{metricphi}) and (\ref{pseudometricphi}) is obtained through
partial derivative in $\Phi$:
\begin{equation}
\breve{g}_{\mu\nu}=\frac{\partial g_{\mu\nu}}{\partial \Phi}
\hspace{1cm}\mbox{and}\hspace{1cm}\breve{g}^{\mu\nu}=
-\frac{\partial g^{\mu\nu}}{\partial \Phi}.
\end{equation}
also one has
$g_{\mu\nu}=\dfrac{\partial^2g_{\mu\nu}}{\partial\Phi^2}$ and
$g^{\mu\nu}=\dfrac{\partial^2g^{\mu\nu}}{\partial\Phi^2}$.

This `adjoint metric field' (\ref{pseudometricphi}) helps us to
simplify the Christoffel symbols. But before, it is necessary to
establish some relationships between ${\bf g}$ and $\breve{\bf
g}$,
\begin{scriptsize}
\begin{eqnarray}
\label{gg0}
g_{\mu\nu}\breve{g}^{\nu\alpha}&=&(e^{\Phi}\eta_{\mu\nu}+\sinh(\Phi)\Upsilon_{\mu\nu})(e^{-\Phi}\eta^{\nu\alpha}+\cosh(\Phi)\Upsilon^{\nu\alpha})\cr
&=&{\delta_{\mu}}^{\alpha}+[e^{\Phi}\cosh(\Phi)+e^{-\Phi}\sinh(\Phi)]{\Upsilon_{\mu}}^{\alpha}\cr
&+&\sinh(\Phi)\cosh(\Phi)\Upsilon_{\mu\nu}\Upsilon^{\nu\alpha},
\end{eqnarray}
\end{scriptsize}
with
$\Upsilon_{\mu\nu}\Upsilon^{\nu\alpha}=-2{\Upsilon_{\mu}}^{\alpha}$
and hyperbolic expressions    $e^{\Phi}=\cosh(\Phi)+\sinh(\Phi)$
and $e^{-\Phi}=\cosh(\Phi)-\sinh(\Phi)$ we have:
\begin{equation}
\label{gg1}
g_{\mu\nu}\breve{g}^{\nu\alpha}={\delta_{\mu}}^{\alpha}+{\Upsilon_{\mu}}^{\alpha},
\end{equation}
and,
\begin{equation}
\label{gg2}
g^{\mu\nu}\breve{g}_{\nu\alpha}={\delta^{\mu}}_{\alpha}+ {\Upsilon^{\mu}}_{\alpha}.
\end{equation}
It is important to observe that:
\begin{equation}
g_{\mu\nu}\breve{g}^{\mu\nu}=g^{\mu\nu}\breve{g}_{\mu\nu}={\bf Tr(\bm{\eta})}+ {\bf Tr(\bm{\Upsilon})}.
\end{equation}
Further we have:
\begin{equation}
\label{gUspsilon1}
g_{\mu\nu}\Upsilon^{\nu\alpha}=(e^{\Phi}\eta_{\mu\nu} +\sinh(\Phi)\Upsilon_{\mu\nu})\Upsilon^{\nu\alpha}=e^{-\Phi}{\Upsilon_\mu}^{\alpha},
\end{equation}
\begin{equation}
\label{gUspsilon2}
g^{\mu\nu}\Upsilon_{\nu\alpha}=e^{\Phi}{\Upsilon^\mu}_{\alpha},
\end{equation}
\begin{equation}
\label{gUspsilon3}
\breve{g}_{\mu\nu}\Upsilon^{\nu\alpha}=-e^{-\Phi}{\Upsilon_\mu}^{\alpha},
\end{equation}
\begin{equation}
\label{gUspsilon4}
\breve{g}^{\mu\nu}\Upsilon_{\nu\alpha}=-e^{\Phi}{\Upsilon^\mu}_{\alpha}.
\end{equation}

Nevertheless for this `adjoint metric field', when there is not
gravitational strength ($\Phi=0$) it is not possible to reorder
Minkowski spacetime. Because if $\Phi=0$ then
$g_{\mu\nu}=\eta_{\mu\nu}+\Upsilon_{\mu\nu}$. Unless all
$\Upsilon_{\mu\nu}$ are vanished, one can not obtain Minkowski
spacetime. Moreover, with $\Upsilon_{\mu\nu}\neq 0 $ a spacetime
$(\cal M,\breve{\bf g})$ has a different topology of physical
spacetimes. For example, if one chooses $\Upsilon_{\mu\nu}$ from
equation (\ref{weak}), the correspondent  line element is
\begin{equation}
d\breve{s}^2=e^{\Phi}c^2dt^2 + e^{-\Phi}(dx^2+dy^2+dz^2),
\end{equation}
that has signature $(++++)$.

\subsection{Christoffel Symbols}

We want a manifold $\cal M$ with Levi-Civita connection, since the
connection coefficient $\bm \Gamma$ is given by
\begin{equation}
\Gamma^{\beta}_{\mu\nu}=\frac{1}{2}g^{\alpha\beta}(\partial_{\mu}g_{\alpha\nu}+\partial_{\nu}g_{\alpha\mu}-\partial_{\alpha}g_{\mu\nu}).
\end{equation}
We have that,
\begin{equation}
\label{pg1}
\partial_{\alpha}g_{\mu\nu}=(e^{\Phi}\eta_{\mu\nu} +\cosh(\Phi)\Upsilon_{\mu\nu})\partial_{\alpha}\Phi+\sinh(\Phi)\partial_{\alpha}\Upsilon_{\mu\nu},
\end{equation}
that in term of (\ref{pseudometricphi}) is
\begin{equation}
\partial_{\alpha}g_{\mu\nu}=\breve{g}_{\mu\nu}\partial_{\alpha}\Phi+\sinh(\Phi)\partial_{\alpha}\Upsilon_{\mu\nu},
\end{equation}
from which we obtain that Christoffel symbols are
\begin{scriptsize}
\begin{eqnarray}
\Gamma^{\beta}_{\mu\nu}&=&\frac{1}{2}g^{\alpha\beta}\Big[\breve{g}_{\alpha\nu}\partial_{\mu}\Phi+\breve{g}_{\alpha\mu}\partial_{\nu}\Phi-\breve{g}_{\mu\nu}\partial_{\alpha}\Phi\cr &+&\sinh(\Phi)(\partial_{\mu}\Upsilon_{\alpha\nu}+\partial_{\nu}\Upsilon_{\alpha\mu}-\partial_{\alpha}\Upsilon_{\mu\nu})\Big]\cr
&=&\frac{1}{2}\left[({\delta^{\beta}}_{\nu}+{\Upsilon^{\beta}}_{\nu})\partial_{\mu}\Phi+({\delta^{\beta}}_{\mu}+{\Upsilon^{\beta}}_{\mu})\partial_{\nu}\Phi-g^{\alpha\beta}\breve{g}_{\mu\nu}\partial_{\alpha}\Phi\right]
\cr &+&\frac{1}{2}\sinh(\Phi)g^{\alpha\beta}(\partial_{\mu}\Upsilon_{\alpha\nu}+\partial_{\nu}\Upsilon_{\alpha\mu}-\partial_{\alpha}\Upsilon_{\mu\nu}).
\end{eqnarray}
\end{scriptsize}
%or
%\begin{scriptsize}
%\begin{eqnarray}
%\label{Cristoffel1}
%\Gamma^{\beta}_{\mu\nu}&=&\frac{1}{2}\left({\delta^{\beta}}_{\nu}{\delta^{\alpha}}_{\mu}+{\delta^{\beta}}_{\mu}{\delta^{\alpha}}_{\nu}+{\delta^{\alpha}}_{\mu}{\Upsilon^{\beta}}_{\nu}+{\delta^{\alpha}}_{\nu}{\Upsilon^{\beta}}_{\mu}-g^{\alpha\beta}\breve{g}_{\mu\nu}\right)\partial_{\alpha}\Phi
%\cr &+&\frac{1}{2}\sinh(\Phi)g^{\alpha\beta}(\partial_{\mu}\Upsilon_{\alpha\nu}+\partial_{\nu}\Upsilon_{\alpha\mu}-\partial_{\alpha}\Upsilon_{\mu\nu})
%\end{eqnarray}
%\end{scriptsize}
Let us consider Christoffel symbols as a combination of two terms: the first is dependent of $\partial_{\alpha}\Phi$
and the second is dependent of $\partial_{\alpha}\bm{\Upsilon}$,
\begin{equation}
\label{Cristoffel1A}
\Gamma^{\beta}_{\mu\nu}=\Gamma^{\beta(1)}_{\mu\nu}+\Gamma^{\beta(2)}_{\mu\nu}
\end{equation}
where,
\begin{scriptsize}
\begin{equation}
\label{Cristoffel2}
\Gamma^{\beta(1)}_{\mu\nu}=\frac{1}{2}\left({\delta^{\beta}}_{\nu}{\delta^{\alpha}}_{\mu}+{\delta^{\beta}}_{\mu}{\delta^{\alpha}}_{\nu}+{\delta^{\alpha}}_{\mu}{\Upsilon^{\beta}}_{\nu}+{\delta^{\alpha}}_{\nu}{\Upsilon^{\beta}}_{\mu}-g^{\alpha\beta}\breve{g}_{\mu\nu}\right)\partial_{\alpha}\Phi
\end{equation}
\end{scriptsize}
and
\begin{equation}
\label{Cristoffel3}
\Gamma^{\beta(2)}_{\mu\nu}=\frac{1}{2}\sinh(\Phi)g^{\alpha\beta}(\partial_{\mu}\Upsilon_{\alpha\nu}+\partial_{\nu}\Upsilon_{\alpha\mu}-\partial_{\alpha}\Upsilon_{\mu\nu}).
\end{equation}
Since $\bm\Gamma$ is not a tensor, it can not have intrinsic geometrical meaning as measures of how much a manifold is curved. Below we shall compute intrinsic objects as Ricci tensor and scalar curvature.

\subsection{Determinant $\bm g$}

From identity \citep{Schutz},
\begin{equation}
\label{ID determinante}
\Gamma^{\nu}_{\mu\nu}=\partial_{\mu}(\ln\,\sqrt{-g}),
\end{equation}
where $g=\det(g_{\mu\nu})$, it is possible to obtain the determinant $g$ by contracting the indices in
(\ref{Cristoffel1A}) with $\beta=\nu$ and the equation (\ref{gg2}):
\begin{small}
\begin{eqnarray}
\label{Cristoffel ID}
\Gamma^{\nu}_{\mu\nu}&=&\frac{1}{2}\Big[{\bf Tr}(\bm{\eta})\partial_{\mu}\Phi+ \partial_{\mu}\Phi+{\bf Tr}(\bm{\Upsilon})\partial_{\mu}\Phi +{\Upsilon^{\alpha}}_{\mu}\partial_{\alpha}\Phi\cr 
& &-  ({\delta^{\alpha}}_{\mu}+{\Upsilon^{\alpha}}_{\mu})\partial_{\alpha}\Phi\Big]
\cr & &+ \frac{1}{2}\sinh(\Phi)g^{\alpha\nu}(\partial_{\mu}\Upsilon_{\alpha\nu}+\partial_{\nu}\Upsilon_{\alpha\mu}-\partial_{\alpha}\Upsilon_{\mu\nu})\cr
&=& \frac{1}{2}\left[{\bf Tr}(\bm{\eta})+{\bf Tr}(\bm{\Upsilon})\right]\partial_{\mu}\Phi+\frac{1}{2}\sinh(\Phi)g^{\alpha\nu}\partial_{\mu}\Upsilon_{\alpha\nu},\nonumber
\end{eqnarray}
\end{small}
we note that the second term with $g^{\alpha\nu}\partial_{\mu}\Upsilon_{\alpha\nu}$ is vanished,
\begin{small}
\begin{eqnarray}
g^{\alpha\nu}\partial_{\mu}\Upsilon_{\alpha\nu}&=&e^{-\Phi}\eta^{\alpha\nu}\partial_{\mu}\Upsilon_{\alpha\nu}-\sinh(\Phi)\Upsilon^{\alpha\nu}\partial_{\mu}\Upsilon_{\alpha\nu}\cr 
&=&e^{-\Phi}\partial_{\mu}{\bf Tr}({\bm\Upsilon})-\sinh(\Phi)[-\partial_{\mu}{\bf Tr}({\bm\Upsilon})]=0,\nonumber
\end{eqnarray}
\end{small}
where we used (\ref{derivada traco2}) and (\ref{Dtraca}), thus,
\begin{eqnarray}
\label{Cristoffel ID2}
\Gamma^{\nu}_{\mu\nu}&=&\frac{1}{2}\left[{\bf Tr}(\bm{\eta})+{\bf Tr}(\bm{\Upsilon})\right]\partial_{\mu}\Phi=\partial_{\mu}(\ln\,\sqrt{-g}).
\end{eqnarray}
and we finally find,
\begin{equation}
\label{ID determinante3}
\sqrt{-g}=\exp\left\{\frac{\Phi}{2}\left[{\bf Tr}(\bm{\eta})+{\bf Tr}(\bm{\Upsilon})\right]\right\}.
\end{equation}
In fact, since $\ln(\det{g_{\mu\nu}})={\bf Tr}(\ln g_{\mu\nu}) $. Also, one should use the identity $dg=g\,\,g^{\mu\nu}dg_{\mu\nu}$ and to compute $\sqrt{-g}$. One can verify the examples of above section in  coordinates $(t,x,y,z)$:

\begin{table}[h]
\begin{scriptsize}
%\caption{One-column table}
\begin{tabular}{@{}ll}
%\hline
%
Yilmaz metric: & $g=-\exp\left\{{\Phi}\left[4+(-6)\right]\right\}=-e^{-2\Phi}$ \\
circularly polarized wave: & $ g=-\exp\left\{{\Phi}\left[4+(-4)\right]\right\}=-1$\\
rotating bodies: & $ g=-\exp\left\{{\Phi}\left[4+(-2)\right]\right\}=-e^{2\Phi}$
%\end{small}
\end{tabular}
\end{scriptsize}
\end{table}

The natural volume element of manifold $\cal M$ with metric tensor (\ref{metricphi}),  $dv =\sqrt{-g}d^4x$, is invariant under coordinate transformation. An action with Lagrangian $\cal L$ in spacetime $({\cal M},{\bf g})$ is given by,
\begin{equation}
\label{action1}
S= \int_{\cal M}{\cal L}\,\, \sqrt{-g}\,\,\,d^4x,
\end{equation}
then we have,
\begin{equation}
\label{action2}
S= \int_{\cal M}{\cal L}\,\, \exp\left\{\frac{\Phi}{2}\left[{\bf Tr}(\bm{\eta})+{\bf Tr}(\bm{\Upsilon})\right]\right\}\,d^4x.
\end{equation}

\section{Newtonian Limit and Gravitational Waves}

\subsection{Newtonian Limit}
In the Newtonian limit one assumes that velocities are small, $\left|\dfrac{v}{c}\right|\ll1,$ that
gravitational potentials are near their Minkowski values,
\footnote{Observe that $g_{\mu\nu}=e^{\Phi}\eta_{\mu\nu}+\sinh(\Phi)\Upsilon_{\mu\nu}\approx (1+\Phi)\eta_{\mu\nu}+\Phi\Upsilon_{\mu\nu}= \eta_{\mu\nu} +\Phi(\eta_{\mu\nu}+\Upsilon_{\mu\nu})=\eta_{\mu\nu}+h_{\mu\nu}$ and \\
$g^{\mu\nu}\approx \eta^{\mu\nu}-\Phi(\eta^{\mu\nu}+\Upsilon^{\mu\nu})= \eta^{\mu\nu}-h^{\mu\nu}$.}
and that pressures or other mechanical stresses are
negligible compared to the energy densities $|P|\ll \rho c^2$.

The description of Einstein's field equation (\ref{field equation}) will use values from Christoffel symbols from equation (\ref{Cristoffel1A}), where this has two terms: $\Gamma^{\beta(1)}_{\mu\nu}$ dependent of $\partial_{\alpha}\Phi$ and  $\Gamma^{\beta(2)}_{\mu\nu}$ dependent of $\partial_{\alpha}\bm\Upsilon$.
In this section we shall verify the Einstein tensor $G_{\mu\nu}=R_{\mu\nu}-\frac{1}{2}g_{\mu\nu}R$, only in the case that $\partial_{\alpha}\bm\Upsilon=0$, then $\Gamma^{\beta(2)}_{\mu\nu}=0$, and in any coordinates system, the Ricci tensor is given by
\begin{equation}
\label{TensorRicci1}
R_{\mu\nu}^{(1)}=\partial_{\alpha}\Gamma^{\alpha(1)}_{\mu\nu} -\Gamma^{\alpha(1)}_{\mu\beta}\Gamma^{\beta(1)}_{\nu\alpha} - \partial_{\nu}\Gamma^{\alpha(1)}_{\mu\alpha}
+\Gamma^{\alpha(1)}_{\alpha\beta}\Gamma^{\beta(1)}_{\mu\nu}.
\end{equation}
Since  $\partial_{\alpha}\bm\Upsilon=0$ the first term of $R_{\mu\nu}^{(1)}$ is given by
\begin{small}
\begin{eqnarray}
\label{Newton 1}
\partial_{\alpha}\Gamma^{\alpha(1)}_{\mu\nu}&=& 
\frac{1}{2}\left(\breve{g}_{\mu\nu}\breve{g}^{\alpha\beta}-g_{\mu\nu}g^{\alpha\beta}\right)\partial_{\alpha}\Phi\partial_{\beta}\Phi
\cr & &+\frac{1}{2}\big(2\partial_{\mu}\partial_{\nu}\Phi+{\Upsilon^{\beta}}_{\nu}\partial_{\mu}\partial_{\beta}\Phi+
{\Upsilon^{\beta}}_{\mu}\partial_{\nu}\partial_{\beta}\Phi\cr
& &-\breve{g}_{\mu\nu}g^{\alpha\beta}\partial_{\alpha}\partial_{\beta}\Phi\big),
\end{eqnarray}
\end{small}
the second term of Ricci tensor is:
\begin{small}
\begin{eqnarray}
\Gamma^{\alpha(1)}_{\mu\beta}\Gamma^{\beta(1)}_{\nu\alpha}&=&\frac{3}{2}\,\,\partial_{\mu}\Phi\,\,\partial_{\nu}\Phi+\frac{1}{2}(\partial_{\mu}\Phi{\Upsilon^{\gamma}}_{\nu}+\partial_{\nu}\Phi{\Upsilon^{\gamma}}_{\mu})\partial_{\gamma}\Phi\cr
& &-\frac{1}{2}g_{\mu\nu}g^{\lambda\gamma}\partial_{\gamma}\Phi\,\,\partial_{\lambda}\Phi
+\frac{1}{2}{\Upsilon^{\lambda}}_{\mu}{\Upsilon^{\gamma}}_{\nu}\partial_{\gamma}\Phi\,\,\partial_{\lambda}\Phi,\nonumber
\end{eqnarray}
\end{small}
with (\ref{Cristoffel ID2}) we can obtain the third term:
\begin{equation}
\label{Newton 3}
\partial_{\nu}\Gamma^{\alpha(1)}_{\mu\alpha}=\frac{1}{2}[{\bf Tr}({\bm \eta})+{\bf Tr}({\bm \Upsilon})]\partial_{\mu}\partial_{\nu}\Phi,
\end{equation}
and the fourth is
%\begin{small}
\begin{eqnarray}
\Gamma^{\alpha(1)}_{\beta\alpha}\Gamma^{\beta(1)}_{\mu\nu}=\frac{1}{4}[{\bf Tr}({\bm \eta})+{\bf Tr}
({\bm \Upsilon})]\big(2\partial_{\mu}\Phi\partial_{\nu}\Phi\cr
+\partial_{\mu}\Phi{\Upsilon^{\beta}}_{\nu}\partial_{\beta}\Phi+
\partial_{\nu}\Phi{\Upsilon^{\beta}}_{\mu}\partial_{\beta}\Phi
-\breve{g}_{\mu\nu}g^{\alpha\beta}\partial_{\alpha}\Phi\partial_{\beta}\Phi \big).
\end{eqnarray}
%\end{small}
For Newtonian limit, we must retain only the terms in the first order in the strength of gravity $\Phi$, or in other words, we have $\partial\Phi\partial\Phi\ll\partial\partial\Phi$ and $\Phi\partial\partial\Phi\ll\partial\partial\Phi$ , with this assumption we have the second term of  (\ref{Newton 1}) and (\ref{Newton 3})   such as,
\begin{eqnarray}
\label{TensorRicci2}
R_{\mu\nu}^{(1)}=\frac{1}{2}\big(2\partial_{\mu}\partial_{\nu}\Phi+{\Upsilon^{\beta}}_{\nu}\partial_{\mu}\partial_{\beta}\Phi
+{\Upsilon^{\beta}}_{\mu}\partial_{\nu}\partial_{\beta}\Phi\cr
-\breve{g}_{\mu\nu}g^{\alpha\beta}\partial_{\alpha}\partial_{\beta}\Phi\big)-
\frac{1}{2}[{\bf Tr}({\bm \eta})+{\bf Tr}({\bm \Upsilon})]\partial_{\mu}\partial_{\nu}\Phi.
\end{eqnarray}
One can choose the simplest $\bm\Upsilon$, and it should be from (\ref{weak}) such as 
${\bf Tr}({\bm \Upsilon})=-6$, where the Ricci tensor is now given by
\begin{scriptsize}
\begin{equation}
R_{\mu\nu}^{(1)}=2\partial_{\mu}\partial_{\nu}\Phi+\frac{1}{2}{\Upsilon^{\beta}}_{\nu}\partial_{\mu}\partial_{\beta}\Phi+\frac{1}{2}{\Upsilon^{\beta}}_{\mu}\partial_{\nu}\partial_{\beta}\Phi-\frac{1}{2}\breve{g}_{\mu\nu}g^{\alpha\beta}\partial_{\alpha}\partial_{\beta}\Phi.
\end{equation}
\end{scriptsize}
The scalar curvature is obtained by further contracting indices
$R^{(1)}=g^{\mu\nu}R_{\mu\nu}^{(1)}$,
\begin{eqnarray}
R^{(1)}= 3g^{\mu\nu}\partial_{\mu}\partial_{\nu}\Phi+g^{\mu\nu}{\Upsilon^{\beta}}_{\nu}\partial_{\mu}\partial_{\beta}\Phi,
\end{eqnarray}
$g^{\mu\nu}=(1-\Phi)\eta^{\mu\nu}-\Phi\Upsilon^{\mu\nu}$ in terms of the first order in $\Phi$, we find that the scalar curvature is
\begin{equation}
R^{(1)}=3\eta^{\mu\nu}\partial_{\mu}\partial_{\nu}\Phi+\eta^{\mu\nu}{\Upsilon^{\beta}}_{\nu}\partial_{\mu}\partial_{\beta}\Phi.
\end{equation}
Since the Newtonian limit approach is in the
background Minkowski spacetime, we find that
\begin{equation}
\label{weak R}
R^{(1)}=3\Box\Phi+\Upsilon^{\beta\mu}\partial_{\mu}\partial_{\beta}\Phi.
\end{equation}
From the field equations (\ref{field equation}) we have
\begin{equation}
\label{field equation2}
R=-\frac{8\pi G}{c^4}T,
\end{equation}
in the situations where Newtonian theory can be applicable
$T\approx{\rho}{c^2}$. Now we can combine equations
(\ref{field equation2}) and (\ref{weak R}) such as
\begin{equation}
\label{weak R2}
3\Box\Phi+\Upsilon^{\beta\mu}\partial_{\mu}\partial_{\beta}\Phi=-\frac{8\pi G}{c^2}\rho.
\end{equation}
We assume that velocities are small, then
\begin{equation}
\Box=\dfrac{\partial^2}{c^2\partial t^2}-\nabla^2\approx -\nabla^2.
\end{equation}
And in particular from
(\ref{weak}) we have $\Upsilon_{\mu\nu}\Rightarrow 2\delta_{ij}$, where $\delta_{ij}$ is the Kronecker delta in   $\mathbb{R}^3$. The field equation (\ref{weak R2}) results in:
\begin{equation}
\label{weak R3}
-3\nabla^2\Phi+2\delta^{ij}\partial_{i}\partial_{j}\Phi=-\frac{8\pi G}{c^2}\rho,
\end{equation}
or
\begin{equation}
\label{weak R4}
\nabla^2\Phi=\frac{8\pi G}{c^2}\rho.
\end{equation}
We can introduce $\Phi=\dfrac{2{\varphi}_{N}}{c^2}$, where ${\varphi}_{N}$ is the Newtonian potential.
Thus we can obtain the Poisson's equation of Newton's law of gravitation
\begin{equation}
\label{weak R5}
\nabla^2\varphi_{N}=4\pi G\rho,
\end{equation}
in this way for a point source with mass $M$ we have
\begin{equation}
\label{potencial de newton}
\varphi_{N}=-\frac{GM}{r}.
\end{equation}
Consequently $\Phi=-\dfrac{2GM}{c^2r}$. In this discussion that follows (\ref{TensorRicci2}) we assumed that $\Upsilon_{\mu\nu}$ is from (\ref{weak}), 
it results that the line element is given by  (\ref{metricphi3}) at thefirst order in $\Phi$.

\subsection{Gravitational Waves}

Gravitational waves are one of the most important physical
phenomena associated with the presence of strong and dynamic
gravitational fields. Though such gravitational radiation has not
yet been detected directly. There is strong indirect evidence for
its existence around the famous binary pulsar PSR 1913+16
\citep{Taylor}, that in 1974 it was discovered by R.A. Hulse and J.H. Taylor \citep {Hulse}, a discovery for which
they were awarded the 1993 Nobel Prize.

Here, for a description of plane gravitational waves we can assume that
$\Phi$ is constant so $\partial_{\alpha}\Phi=0$, implying in $\Gamma^{\beta(1)}_{\mu\nu}=0$ in
(\ref{Cristoffel1A}), thus we have $\Gamma^{\beta}_{\mu\nu}=\Gamma^{\beta(2)}_{\mu\nu}$ given by equation (\ref{Cristoffel3}):
\begin{equation}
\label{Cristoffel5}
\Gamma^{\beta(2)}_{\mu\nu}=\frac{1}{2}\sinh(\Phi)g^{\alpha\beta}(\partial_{\mu}\Upsilon_{\alpha\nu}+\partial_{\nu}\Upsilon_{\alpha\mu}-\partial_{\alpha}\Upsilon_{\mu\nu}).
\end{equation}
One can obtain the field equation writing the Ricci tensor in terms of Christoffel symbols above,
\begin{equation}
\label{TensorRicci2a}
R_{\mu\nu}^{(2)}=\partial_{\alpha}\Gamma^{\alpha(2)}_{\mu\nu} -\Gamma^{\alpha(2)}_{\mu\beta}\Gamma^{\beta(2)}_{\nu\alpha} - \partial_{\nu}\Gamma^{\alpha(2)}_{\mu\alpha}
+\Gamma^{\alpha(2)}_{\alpha\beta}\Gamma^{\beta(2)}_{\mu\nu}.
\end{equation}
From equation (\ref{Cristoffel ID2}) we have \\$\Gamma^{\alpha}_{\alpha\nu}=\dfrac{1}{2}[{\bf Tr}({\bm \eta}) +{\bf Tr}({\bm\Upsilon})]\partial_{\nu}\Phi=0$, such as,
\begin{equation}
\label{TensorRicci2b}
R_{\mu\nu}^{(2)}=\partial_{\alpha}\Gamma^{\alpha(2)}_{\mu\nu} -\Gamma^{\alpha(2)}_{\mu\beta}\Gamma^{\beta(2)}_{\nu\alpha},
\end{equation}
As shown in appendix, the 
The above Ricci tensor is given by,
\begin{small}
\begin{eqnarray}
\label{TensorRicci2c}
R_{\mu\nu}^{(2)}=\frac{1}{2}\sinh^2(\Phi)\,\partial_\alpha
\Upsilon^{\alpha\beta}(\partial_\mu\Upsilon_{\beta\nu}+\partial_\nu\Upsilon_{
\beta\mu}-\partial_{\beta}\Upsilon_{\mu\nu})& &\cr
+\frac{1}{2}\sinh(\Phi)\,
g^{\alpha\beta}(\partial_\alpha\partial_\mu\Upsilon_{\beta\nu}
+\partial_\alpha\partial_\nu\Upsilon_{\beta\mu}-\partial_\alpha\partial_{\beta}
\Upsilon_{\mu\nu})& &\cr
-\frac{1}{4}\sinh^2(\Phi)[g^{\alpha\gamma}g^{\beta\lambda}\partial_\mu\Upsilon_{
\gamma\beta}\partial_\nu\Upsilon_{\alpha\lambda}& &\cr
+2(g^{\alpha\gamma}g^{\beta\lambda}-g^{\alpha\beta}g^{\gamma\lambda}
)\partial_\beta\Upsilon_{\gamma\mu}\partial_\alpha\Upsilon_{\nu\lambda}].& &
\end{eqnarray}
\end{small}
The study of gravitational waves involves essentially the approximation 
of Einstein's weak field equation, from these results one can obtain wave equation with low amplitudes,  $\Phi\ll 1$. This can only be brought in the first
order of the gravitational potential to find,
\begin{equation}
\label{TensorRicci2d}
R_{\mu\nu}^{(2)}=\frac{1}{2}\sinh(\Phi)\, g^{\alpha\beta}(\partial_\alpha\partial_\mu\Upsilon_{\beta\nu}+\partial_\alpha\partial_\nu\Upsilon_{\beta\mu}-\partial_\alpha\partial_{\beta}\Upsilon_{\mu\nu}),
\end{equation}
with $\sinh(\Phi)\, g^{\alpha\beta}\approx \Phi[(1-\Phi)\eta^{\alpha\beta}-\Phi\Upsilon^{\alpha\beta}]=\Phi\eta^{\alpha\beta}$, one can obtain the Ricci tensor in Minkowski background spacetime, which this yields,
\begin{equation}
\label{TensorRicci2e}
R_{\mu\nu}^{(2)}=\frac{1}{2}{\Phi}\,(\partial^\beta\partial_\mu\Upsilon_{\beta\nu}+\partial^\beta\partial_\nu\Upsilon_{\beta\mu}-\Box\Upsilon_{\mu\nu}),
\end{equation}
The gauge choice $\partial^\beta\Upsilon_{\beta\nu}=0 $,
simplifies Ricci tensor in:
\begin{equation}
\label{TensorRicci2f}
R_{\mu\nu}^{(2)}=-\frac{1}{2}{\Phi}\,\Box\Upsilon_{\mu\nu}.
\end{equation}
One should observe that scalar curvature 
\begin{equation}
R^{(2)}=g^{\mu\nu}R_{\mu\nu}^{(2)}\nonumber 
\end{equation}
is vanished for the first order in $\Phi$, because 
\begin{equation}
R^{(2)}=\dfrac{1}{2}\eta^{\mu\nu}{\Phi}\,\Box\Upsilon_{\mu\nu}=\dfrac{1}{2}\Box{\bf Tr}({\bm \Upsilon})=0, \nonumber
\end{equation}
since the trace of tensor $\bm\Upsilon$ is a constant number, in accordance with equation (\ref{Dtraca}). Then, for the field equation (\ref{field equation}) the gravitational wave equation obtained in the vacuum is only,
\begin{equation}
\label{wave1}
\Box\Upsilon_{\mu\nu}=0.
\end{equation}
It is necessary that the tensor
$\Upsilon_{\mu\nu}$ satisfies the condition (\ref{ID}) and the above wave equation.
Its shape should be, for example, the tensor (\ref{wavephi}) with $\zeta=\kappa_{\mu}x^{\mu}=\omega t-kz$,
a solution for gravitational plane waves with circular polarization traveling in the
$z$ direction, with amplitude $\Phi$.

\section{Conclusion}

This paper deals with a tensorial structure that assumes a
(quasi-)idempotent feature to be able to improve at least the
linear tensorial template of some tensor metric fields. It is
clear that Einstein's field equations are non-linear, however,
with these  (quasi-)idempotent tensorial structure, without
quadratic tensorial values, the non-linearity becomes more
moderate although there is a price to pay. The part that
carries  the dynamical information, the strength of gravity is
tied to the tensorial structure by exponential functions. In this
approach the metric field can be characterized by a background
spacetime conformally flat affected by a disturbance. We have approached some examples
in this tensorial structure that results in exponential metric fields, we
can point out as the main exponential metric obtained in this paper which
has been extensively explored: the Yilmaz exponential metric
\citep{Yilmaz1,Yilmaz2,Yilmaz3,Yilmaz4,Yilmaz5,Yilmaz6,Yilmaz7,Clapp,Robertson1,Robertson2,Ibison}.
H. Yilmaz has argued that in his theory, the gravitational field
can be quantized via Feynman's method \citep{Yilmaz9,Alley}.
Further, it has been found that the quantized theory is finite.
Incidentally in the exponential metric fields approached in this
work just as in the Yilmaz theory there are no black holes in the
sense of event horizons, but there can be stellar collapse
\citep{Robertson1,Robertson2}. However, there are no point
singularities.

Interesting results obtained in this work from exponential metric fields
are: circularly polarized wave; rotating bodies that in the
first order is a deformation of Kerr metric and also we have a deformed static spherically symmetric
spacetime. Many discussions around massive stellar objects have suggested,
for example, that Kerr metric should  be slightly deviated from Kerr.
The possibility of discovering a non-Kerr
object should be taken into account when
constructing waveform templates for LISA's data analysis tools \citep{Glampedakis,White}.
The technological development is ripe enough so much so in the years to come
we might be able to test the second
order relativist-gravity effects and may lead to answers to some important questions of gravity.

In this work, we have obtained a simple and general expression for the volume element of a manifold 
in coordinates $(t,x,y,z)$ given in terms of strength of gravity and of traces of 
tensors $\bm \eta$ and $\bm \Upsilon$. It is possible that an analysis of any Lagrangian of field 
interacting with gravity will become easer.  An interesting observation is the spacetime of
circularly polarized plane wave, in this spacetime the volume element $\sqrt{-g}\,\,d^4x$ is the same
of Minkowski spacetime, in this sense this  gravitational radiation obtained from exponential metric field
does not modify the volume element of background Minkowski spacetime where this plane wave 
travels onto. 
%%%%%%%%%%%%%%%%%%%%%%%%%%%%%%%%%%%%%%%%%%%%%%%%%%%%%%%%%%%%%
Moreover, it was purposed and verified the Newtonian limit as
solution for Einstein's equation, since we can assume that the
trace of stress-energy tensor is $T\approx \rho c^2$. Other
important solution of Einstein's equation analysed in this paper
was the plane gravitational wave for the empty space since we have
considered the vanished stress-energy tensor to the first order in $\Phi$.
Both solved Einstein's equations for Newtonian limit and plane gravitational wave
propagating in the vacuum are cases that the strength of gravity is small, $\Phi\ll 1$.
%%%%%%%%%
%Should it be possible that the solution for small $\Phi$ can be valid for any order?
%In affirmative case, it should be possible in many ways, to solve Einstein's equations
%in low energy, with $\Phi\ll 1$ only in the first order, and with 
%$g_{\mu\nu}=e^{\Phi}\eta_{\mu\nu}+\sinh(\Phi)\Upsilon_{\mu\nu}$ to obtain the metric field in any order.
We have analysed  the Newtonian limit  in the case that $\partial_{\alpha}\bm\Upsilon=0$, and analysed
the plane gravitational wave considering  the strength of gravity as a constant term, thus we had 
two independent Ricci tensors, $R_{\mu\nu}^{(1)}$ which $\partial_{\alpha}\bm\Upsilon=0$ (for Newtonian limit)
and $R_{\mu\nu}^{(2)}$ which $\partial_{\alpha}\Phi=0$ (for plane gravitational wave). In a forthcoming work, 
an analysis of Einstein's equations with both non-vanished $\partial_{\alpha}\bm\Upsilon$ 
and $\partial_{\alpha}\Phi$, will be considered.

It is missing a discussion about quantities of physical interest
in the solutions of Einstein's equations which describe the
exterior and interior gravitational field. Yilmaz has argued the
existence of the matter part in the right-hand side of the field
equations correspondent to field energy in the exterior. This
paper lacks a discussion about the interior and the exterior field
energies denoted by a total stress-energy tensor. An analysis
about the total stress-energy tensor will be the object 
of a forthcoming study, where the physical consequences of terms of
deformity in Kerr and Schwarzschild solutions could be analysed.
%%%%%%%%%%%%%%%%%%%%%%%%%
%%%%%%%%%%%%%%%%%%%%%%%%%
%%%%%%%%%%%%%%%%%%%%%%%%%
%%%%%%%%%%%%%%%%%%%%%%%%%
%%%%%%%%%%%%%%%%%%%%%%%%%
%%%%%%%%%%%%%%%%%%%%%%%%%
%%%%%%%%%%%%%%%%%%%%%%%%%
%%%%%%%%%%%%%%%%%%%%%%%%%

We know that the dark energy and the dark matter problems are challenges to modern astrophysics and cosmology;
as a typical example, we could mention the galactic rotation curves of spiral galaxies, that probably, indicates the possible 
failure of both Newtonian gravity and General Relativity
on galactic and intergalactic scales. To explain astrophysical and cosmological problems with arguments against 
dark energy and dark matter many works have been devoted to the possibility that 
the Einstein-Hilbert Lagrangian, linear in the Ricci scalar R, should be generalized.
In this sense, the choice of a generic function $f(R)$
can be derived by matching the data and the requirement that no
exotic ingredient have to be added 
\citep{Allemandi,Barrow,Capozziello1,Capozziello2,Capozziello3,Carroll1,Carroll2,Faraoni,Flanagan,Koivisto,Nojiri1,Nojiri2,Nojiri3,Sotiriou}. 
This class of theories when linearized exhibits others polarization modes for
the gravitational waves, 
of which two correspond to the massless graviton and others such as massive scalar and ghost modes in $f(R)$ gravity 
\citep{Bellucci,Bogdanos}. In this way, analyses in any order to $f(R)$ gravity with `exponential metrics' proposed in 
the present work could give a positive contribution to the debate of astrophysical and cosmological questions. 
%%%%%%%%%%%%%%%%%%%%%%%%%
%%%%%%%%%%%%%%%%%%%%%%%%%
%%%%%%%%%%%%%%%%%%%%%%%%%
%%%%%%%%%%%%%%%%%%%%%%%%%
%%%%%%%%%%%%%%%%%%%%%%%%%
%%%%%%%%%%%%%%%%%%%%%%%%%

%%%%%%%%%%%%%%%%%%%%%%%%%
%%%%%%%%%%%%%%%%%%%%%%%%%

\acknowledgments
It is a pleasure to acknowledge many stimulating and helpful discussions with my
colleagues and friends Andr\'e L. A. Penna and Caio M. M. Polito.

\appendix

\section{Details of Calculation for Rotating Bodies}
\subsection{Multiplicative Properties of $\bm\Upsilon$ in Boyer-Lindquist
coordinates}
The metric tensor $\eta_{\mu\nu}$ of Minkowski flat spacetime in Boyer-Lindquist
coordinates is given by:
\begin{equation}
\eta_{\mu\nu}=
 \left[ \begin {array}{cccc} 1&0&0&0\\\noalign{\medskip}0&-{\frac {{r}
^{2}+{a}^{2} \left( \cos \left( \theta \right)  \right) ^{2}}{{r}^{2}+
{a}^{2}}}&0&0\\\noalign{\medskip}0&0&-{r}^{2}-{a}^{2} \left( \cos
 \left( \theta \right)  \right) ^{2}&0\\\noalign{\medskip}0&0&0&-
 \left( {r}^{2}+{a}^{2} \right)  \left( \sin \left( \theta \right)
 \right) ^{2}\end {array} \right],
\end{equation}
while the (quasi-)idempotent tensor $\Upsilon_{\mu\nu}$ from (\ref{UKerr}) in
the same coordinates is:
\begin{equation}
\Upsilon_{\mu\nu}= (-2)
\left[ \begin {array}{cccc}  \left( \cosh \left( \Lambda \right)
 \right) ^{2}&0&0&\sinh \left( \Lambda \right) \cosh \left( \Lambda
 \right) \sqrt {{r}^{2}+{a}^{2}}\sin \left( \theta \right)
\\\noalign{\medskip}0&0&0&0\\\noalign{\medskip}0&0&0&0
\\\noalign{\medskip}\sinh \left( \Lambda \right) \cosh \left( \Lambda
 \right) \sqrt {{r}^{2}+{a}^{2}}\sin \left( \theta \right) &0&0&
 \left( \sinh \left( \Lambda \right)  \right) ^{2} \left( {r}^{2}+{a}^
{2} \right)  \left( \sin \left( \theta \right)  \right) ^{2}
\end {array} \right].
\end{equation}
Let us verify that above tensor $\Upsilon_{\mu\nu} $ satisfies the algebraic
relation (\ref{ID}). We begin with the inverse metric tensor of Minkowski space:
\begin{equation}
\eta^{\mu\nu}=
 \left[ \begin {array}{cccc} 1&0&0&0\\\noalign{\medskip}0&-{\frac {{r}
^{2}+{a}^{2}}{{r}^{2}+{a}^{2} \left( \cos \left( \theta \right)
 \right) ^{2}}}&0&0\\\noalign{\medskip}0&0&- \left( {r}^{2}+{a}^{2}
 \left( \cos \left( \theta \right)  \right) ^{2} \right) ^{-1}&0
\\\noalign{\medskip}0&0&0&-{\frac {1}{ \left( {r}^{2}+{a}^{2} \right)
 \left( \sin \left( \theta \right)  \right) ^{2}}}\end {array}
 \right],
\end{equation}
such as ${\Upsilon^{\mu}}_{\nu}$ e ${\Upsilon_{\mu}}^{\nu}$ are calculated with
contracting indices,
\begin{equation}
{\Upsilon^{\mu}}_{\nu}=\eta^{\mu\alpha}\Upsilon_{\alpha\nu}=(-2)
\left[ \begin {array}{cccc}  \left( \cosh \left( \Lambda \right)
 \right) ^{2}&0&0&\sinh \left( \Lambda \right) \cosh \left( \Lambda
 \right) \sqrt {{r}^{2}+{a}^{2}}\sin \left( \theta \right)
\\\noalign{\medskip}0&0&0&0\\\noalign{\medskip}0&0&0&0
\\\noalign{\medskip}-{\frac {\sinh \left( \Lambda \right) \cosh
 \left( \Lambda \right) }{\sqrt {{r}^{2}+{a}^{2}}\sin \left( \theta
 \right) }}&0&0&- \left( \sinh \left( \Lambda \right)  \right) ^{2}
\end {array} \right]
\end{equation}
and
\begin{equation}
\label{apendice5}
{\Upsilon_{\mu}}^{\nu}=\Upsilon_{\mu\alpha}\eta^{\alpha\nu}=(-2)
\left[ \begin {array}{cccc}  \left( \cosh \left( \Lambda \right)
 \right) ^{2}&0&0&-{\frac {\sinh \left( \Lambda \right) \cosh \left(
\Lambda \right) }{\sqrt {{r}^{2}+{a}^{2}}\sin \left( \theta \right) }}
\\\noalign{\medskip}0&0&0&0\\\noalign{\medskip}0&0&0&0
\\\noalign{\medskip}\sinh \left( \Lambda \right) \cosh \left( \Lambda
 \right) \sqrt {{r}^{2}+{a}^{2}}\sin \left( \theta \right) &0&0&-
 \left( \sinh \left( \Lambda \right)  \right) ^{2}\end {array}
 \right].
\end{equation}
Notice that it follows  the contravariant tensor $\Upsilon^{\mu\nu}$ given by:
\begin{equation}
\label{apendice6}
\Upsilon^{\mu\nu}=\eta^{\mu\alpha}\Upsilon_{\alpha\beta}\eta^{\beta\nu}=(-2)
\left[ \begin {array}{cccc}  \left( \cosh \left( \Lambda \right)
 \right) ^{2}&0&0&-{\frac {\sinh \left( \Lambda \right) \cosh \left(
\Lambda \right) }{\sqrt {{r}^{2}+{a}^{2}}\sin \left( \theta \right) }}
\\\noalign{\medskip}0&0&0&0\\\noalign{\medskip}0&0&0&0
\\\noalign{\medskip}-{\frac {\sinh \left( \Lambda \right) \cosh
 \left( \Lambda \right) }{\sqrt {{r}^{2}+{a}^{2}}\sin \left( \theta
 \right) }}&0&0&{\frac { \left( \sinh \left( \Lambda \right)  \right)
^{2}}{ \left( {r}^{2}+{a}^{2} \right)  \left( \sin \left( \theta
 \right)  \right) ^{2}}}\end {array} \right].
\end{equation}
Finally one can obtain that
$\Upsilon_{\mu\nu}\Upsilon^{\nu\alpha}=-2{\Upsilon_{\mu}}^{\alpha}$, this is
proved by observing that,
\begin{equation}
\label{apendice7}
\Upsilon_{\mu\nu}\Upsilon^{\nu\alpha}=(-2)\cdot(-2)
 \left[ \begin {array}{cccc}  \left( \cosh \left( \Lambda \right)
 \right) ^{2}&0&0&-{\frac {\sinh \left( \Lambda \right) \cosh \left(
\Lambda \right) }{\sqrt {{r}^{2}+{a}^{2}}\sin \left( \theta \right) }}
\\\noalign{\medskip}0&0&0&0\\\noalign{\medskip}0&0&0&0
\\\noalign{\medskip}\sinh \left( \Lambda \right) \cosh \left( \Lambda
 \right) \sqrt {{r}^{2}+{a}^{2}}\sin \left( \theta \right) &0&0&-
 \left( \sinh \left( \Lambda \right)  \right) ^{2}\end {array}
 \right] = -2{\Upsilon_{\mu}}^{\alpha}.
\end{equation}
It is easy to see that the above tensor is the same of equation
(\ref{apendice5})  as claimed.

\subsection{Calculation of components}
Let us compute the components of metric tensor
$g_{\mu\nu}=e^{\Phi}\eta_{\mu\nu}+\sinh(\Phi)\Upsilon_{\mu\nu}$ for rotating
bodies. First we may calculate
$g_{00}=e^{\Phi}\eta_{00}+\sinh(\Phi)\Upsilon_{00}\approx
(1+{\Phi})\eta_{00}+{\Phi}\Upsilon_{00}$,
\begin{eqnarray}
g_{00}&=&1+{\Phi}-2{\Phi}\cosh^2{\Lambda}=1+ \frac{Mr}{(r^2+a^2)} -2\cdot
\frac{Mr}{(r^2+a^2)}\cdot \frac{{r^2+a^2}}{\rho^2},
\end{eqnarray}
since $\Phi=\dfrac{Mr}{(r^2+a^2)}$ from (\ref{Phi RB}), and also
$\cosh{\Lambda}=\dfrac{\sqrt{r^2+a^2}}{\rho}$ from (\ref{definicao shL}),
\begin{eqnarray}
g_{00}  &=& 1-
\frac{2Mr}{\rho^2}+\frac{Mr}{(r^2+a^2)}=\frac{\rho^2-2Mr}{\rho^2}+\frac{Mr}{
(r^2+a^2)},
\end{eqnarray}
 with $ \rho=r^2+a^2\cos^2\theta$ then,
\begin{eqnarray}
g_{00} &=& \frac{r^2+a^2\cos^2\theta-2Mr}{\rho^2}+\frac{Mr}{(r^2+a^2)}=
\frac{r^2+a^2(1-\sin^2\theta)-2Mr}{\rho^2}+\frac{Mr}{(r^2+a^2)}\cr
      &=& \frac{r^2+a^2-2Mr-a^2\sin^2\theta}{\rho^2}+\frac{Mr}{(r^2+a^2)}\cr
      &=&\frac{\Delta-a^2\sin^2\theta}{\rho^2}+\frac{Mr}{(r^2+a^2)},
\end{eqnarray}
where  $\Delta=r^2-2Mr-a^2$

Calculation of $g_{03}$:
\begin{eqnarray}
g_{03}&=&-2\sinh(\Phi)\sinh \Lambda \cosh \Lambda \sqrt
{{r}^{2}+{a}^{2}}\sin\theta \cr            
&=&-2\left(\frac{Mr}{r^2+a^2}\right)\left(-\frac{a\sin\theta}{\rho}
\right)\left(\frac{\sqrt{r^2+a^2}}{\rho}\right)\sqrt{r^2+a^2}\sin{\theta}=\frac{
2Mra\sin^2{\theta}}{\rho^2},
\end{eqnarray}
again we used the definition (\ref{definicao shL}).

Calculation of $g_{22}$:
\begin{eqnarray}
g_{22}&=&-e^{\Phi}\rho^2=-\left(1+\frac{Mr}{(r^2+a^2)}
\right)\rho^2=-\rho^2-\frac{Mr\rho^2}{r^2+a^2}
\end{eqnarray}

Calculation of $g_{33}$:
\begin{eqnarray}
g_{33}&=&-e^{\Phi}(r^2+a^2)\sin^2{\theta}-2\sinh(\Phi)\sinh^2{\Lambda}
(r^2+a^2)\sin^2{\theta}\cr
      &=&-\left(1+ \frac{Mr}{(r^2+a^2)}\right)(r^2+a^2)\sin^2{\theta} -
2\left(\frac{Mr}{r^2+a^2}\right)\left(\frac{a^2\sin^2\theta}{\rho^2}
\right)(r^2+a^2)\sin^2{\theta}\cr    
&=&-(r^2+a^2)\sin^2{\theta}-\frac{2Mra^2\sin^4\theta}{\rho^2}-Mr\sin^2\theta\cr 
   &=&-\frac{\rho^2(r^2+a^2)\sin^2{\theta}+2Mra^2\sin^4\theta}{\rho^2}
-Mr\sin^2\theta\cr     
&=&-\frac{(r^2+a^2\cos^2\theta)(r^2+a^2)\sin^2{\theta}+2Mra^2\sin^4\theta}{
\rho^2}-Mr\sin^2\theta\cr     
&=&-\frac{(r^2+a^2-a^2\sin^2\theta)(r^2+a^2)\sin^2{\theta}+2Mra^2\sin^4\theta}{
\rho^2}-Mr\sin^2\theta\cr     
&=&-\frac{\left[(r^2+a^2)-a^2\sin^2\theta\right](r^2+a^2)\sin^2{\theta}
+2Mra^2\sin^4\theta}{\rho^2}-Mr\sin^2\theta\cr     
&=&\frac{-(r^2+a^2)^2\sin^2\theta+(r^2-2Mr+a^2)a^2\sin^4\theta}{\rho^2}
-Mr\sin^2\theta\cr
      &=&\frac{-\sin^2\theta}{\rho^2}\left[\left(r^2+a^2\right)^2-\Delta
a^2\sin^2\theta\right] - Mr\sin^2\theta
\end{eqnarray}

Calculation of $g_{11}$:
\begin{equation}
g_{11}=-e^{\Phi}\frac{\rho^2}{r^2+a^2},
\end{equation}
we have that $\Delta=r^2-2Mr-a^2$ can be given by:
\begin{eqnarray}
\Delta &=& r^2-2Mr-a^2=(r^2+ a^2)-2Mr\cr
\Delta &=&(r^2+ a^2)\left(1-\frac{2Mr}{(r^2+a^2)}\right),
\end{eqnarray}
which implies that
\begin{equation}
\frac{1}{r^2+a^2}=\frac{1}{\Delta}\left(1-\frac{2Mr}{(r^2+a^2)}\right).
\end{equation}
Now, the component  $g_{11}$ is given by:
\begin{eqnarray}
g_{11}&=&-e^{\Phi}\frac{\rho^2}{r^2+a^2}=-\left(1+
\frac{Mr}{(r^2+a^2)}\right)\frac{\rho^2}{\Delta}\left(1-\frac{2Mr}{(r^2+a^2)}
\right).
\end{eqnarray}
Hence we have that $ \dfrac{Mr}{r^2+a^2}\ll 1 $, the component $g_{11}$ 
is given just in the first order:
\begin{eqnarray}
g_{11}&=& -\frac{\rho^2}{\Delta}\left(1-\frac{Mr}{(r^2+a^2)}\right)\cr
g_{11}&=& -\frac{\rho^2}{\Delta}+ \frac{Mr\rho^2}{\Delta(r^2+a^2)}.
\end{eqnarray}
%-------------------------------------------------------------------------------
%-------------------------------------------------------------------------------
%-------------------------------------------------------------------------------
%-------------------------------------------------------------------------------
%-------------------------------------------------------------------------------

\section{Calculation of Ricci tensor $ R^{(2)}_{\mu\nu}$ }
The Ricci tensor from section VI, used to evaluated gravitational waves, is
given by:
\begin{equation}
R_{\mu\nu}^{(2)}=\partial_{\alpha}\Gamma^{\alpha(2)}_{\mu\nu}
-\Gamma^{\alpha(2)}_{\mu\beta}\Gamma^{\beta(2)}_{\nu\alpha} -
\partial_{\nu}\Gamma^{\alpha(2)}_{\mu\alpha}
+\Gamma^{\alpha(2)}_{\alpha\beta}\Gamma^{\beta(2)}_{\mu\nu}.
\end{equation}
Let $\Phi$ be constant. Then $\Gamma^{\alpha}_{\alpha\nu}=\dfrac{1}{2}[{\bf
Tr}({\bm \eta}) +{\bf Tr}({\bm\Upsilon})]\partial_{\nu}\Phi$ are vanished such
that
\begin{equation}
\label{TensorRicciApp}
R_{\mu\nu}^{(2)}=\partial_{\alpha}\Gamma^{\alpha(2)}_{\mu\nu}
-\Gamma^{\alpha(2)}_{\mu\beta}\Gamma^{\beta(2)}_{\nu\alpha}.
\end{equation}
Let us calculate the first term:
\begin{equation}
\partial_{\alpha}\Gamma^{\alpha(2)}_{\mu\nu}=
\frac{1}{2}\sinh(\Phi)\,\partial_\alpha
g^{\alpha\beta}(\partial_\mu\Upsilon_{\beta\nu}+\partial_\nu\Upsilon_{\beta\mu}
-\partial_{\beta}\Upsilon_{\mu\nu})+
\frac{1}{2}\sinh(\Phi)\,
g^{\alpha\beta}(\partial_\alpha\partial_\mu\Upsilon_{\beta\nu}
+\partial_\alpha\partial_\nu\Upsilon_{\beta\mu}-\partial_\alpha\partial_{\beta}
\Upsilon_{\mu\nu}).
\end{equation}
If  $\Phi$ is constant, then we have $\partial_\alpha
g^{\alpha\beta}=\sinh(\Phi)\partial_\alpha \Upsilon^{\alpha\beta}$, such that,
\begin{equation}
\partial_{\alpha}\Gamma^{\alpha(2)}_{\mu\nu}=
\frac{1}{2}\sinh^2(\Phi)\,\partial_\alpha
\Upsilon^{\alpha\beta}(\partial_\mu\Upsilon_{\beta\nu}+\partial_\nu\Upsilon_{
\beta\mu}-\partial_{\beta}\Upsilon_{\mu\nu})+
\frac{1}{2}\sinh(\Phi)\,
g^{\alpha\beta}(\partial_\alpha\partial_\mu\Upsilon_{\beta\nu}
+\partial_\alpha\partial_\nu\Upsilon_{\beta\mu}-\partial_\alpha\partial_{\beta}
\Upsilon_{\mu\nu}).
\end{equation}
Let us calculate the second term from Ricci tensor (\ref{TensorRicciApp}):
\begin{eqnarray}
\Gamma^{\alpha(2)}_{\mu\beta}\Gamma^{\beta(2)}_{\nu\alpha}&=&\frac{1}{2}
\sinh(\Phi)g^{\alpha\gamma}(\partial_\mu\Upsilon_{\gamma\beta}
+\partial_\beta\Upsilon_{\gamma\mu}-\partial_{\gamma}\Upsilon_{\mu\beta})\frac{1
}{2}\sinh(\Phi)g^{\beta\lambda}(\partial_\alpha\Upsilon_{\nu\lambda}
+\partial_\nu\Upsilon_{\alpha\lambda}-\partial_{\lambda}\Upsilon_{\alpha\nu})\cr
&=&\frac{1}{4}\sinh^2(\Phi)g^{\alpha\gamma}g^{\beta\lambda}[
(\partial_\mu\Upsilon_{\gamma\beta}\partial_\alpha\Upsilon_{\nu\lambda}-
\partial_\mu\Upsilon_{\gamma\beta}\partial_{\lambda}\Upsilon_{\alpha\nu}) +
(\partial_\beta\Upsilon_{\gamma\mu}\partial_\alpha\Upsilon_{\nu\lambda}
+\partial_{\gamma}\Upsilon_{\mu\beta}\partial_{\lambda}\Upsilon_{\alpha\nu})\cr
&
&+(\partial_\beta\Upsilon_{\gamma\mu}\partial_{\nu}\Upsilon_{\alpha\lambda}
-\partial_{\gamma}\Upsilon_{\mu\beta}\partial_\nu\Upsilon_{\alpha\lambda}
)-(\partial_\beta\Upsilon_{\gamma\mu}\partial_{\lambda}\Upsilon_{\alpha\nu}
+\partial_{\gamma}\Upsilon_{\mu\beta}\partial_\alpha\Upsilon_{\nu\lambda}) +
\partial_\mu\Upsilon_{\gamma\beta}\partial_\nu\Upsilon_{\alpha\lambda}]
.\nonumber
\end{eqnarray}
The contracting indices $\alpha, \beta, \gamma$ and $\lambda$ can be manipulated
as following: $\alpha\leftrightarrow\lambda$ and $\beta\leftrightarrow\gamma$ in
the second term of each above parentesis, that result in
\begin{equation}
\Gamma^{\alpha(2)}_{\mu\beta}\Gamma^{\beta(2)}_{\nu\alpha}=\frac{1}{4}
\sinh^2(\Phi)g^{\alpha\gamma}g^{\beta\lambda}[\,2\partial_\beta\Upsilon_{
\gamma\mu}\partial_\alpha\Upsilon_{\nu\lambda}-2\partial_{\gamma}\Upsilon_{
\mu\beta}\partial_\alpha\Upsilon_{\nu\lambda}+\partial_\mu\Upsilon_{\gamma\beta}
\partial_\nu\Upsilon_{\alpha\lambda}]
\end{equation}
or
\begin{equation}
\Gamma^{\alpha(2)}_{\mu\beta}\Gamma^{\beta(2)}_{\nu\alpha}=\frac{1}{4}
\sinh^2(\Phi)[g^{\alpha\gamma}g^{\beta\lambda}\partial_\mu\Upsilon_{\gamma\beta}
\partial_\nu\Upsilon_{\alpha\lambda}+2(g^{\alpha\gamma}g^{\beta\lambda}-g^{
\alpha\beta}g^{\gamma\lambda})\partial_\beta\Upsilon_{\gamma\mu}
\partial_\alpha\Upsilon_{\nu\lambda}].
\end{equation}
Finally we have the Ricci tensor given by:
\begin{eqnarray}
R_{\mu\nu}^{(2)}&=&\frac{1}{2}\sinh^2(\Phi)\,\partial_\alpha
\Upsilon^{\alpha\beta}(\partial_\mu\Upsilon_{\beta\nu}+\partial_\nu\Upsilon_{
\beta\mu}-\partial_{\beta}\Upsilon_{\mu\nu})\cr 
& &+\frac{1}{2}\sinh(\Phi)\,
g^{\alpha\beta}(\partial_\alpha\partial_\mu\Upsilon_{\beta\nu}
+\partial_\alpha\partial_\nu\Upsilon_{\beta\mu}-\partial_\alpha\partial_{\beta}
\Upsilon_{\mu\nu})\cr
&
&-\frac{1}{4}\sinh^2(\Phi)[g^{\alpha\gamma}g^{\beta\lambda}\partial_\mu\Upsilon_
{\gamma\beta}\partial_\nu\Upsilon_{\alpha\lambda}+2(g^{\alpha\gamma}g^{
\beta\lambda}-g^{\alpha\beta}g^{\gamma\lambda})\partial_\beta\Upsilon_{\gamma\mu
}\partial_\alpha\Upsilon_{\nu\lambda}].
\end{eqnarray}

\nocite{*}
\bibliographystyle{spr-mp-nameyear-cnd}
\bibliography{myref}
\bibliography{biblio-u1}

%\end{document}

\end{document}